\documentclass[journal,10pt]{IEEEtran}
\usepackage{graphicx}   
\usepackage{bm}
\usepackage{amsmath}
\usepackage{stmaryrd}
\allowdisplaybreaks
\usepackage{amsfonts}
\usepackage{color}
\usepackage{ulem}
\usepackage{stfloats}
\usepackage{subfigure}
\usepackage{float}
\usepackage{amssymb}
\usepackage{amsthm}
\usepackage{algpseudocode}  
\usepackage{algorithmicx,algorithm}
\usepackage{cite}
\usepackage[colorlinks,linkcolor=blue]{hyperref}

\usepackage{caption}
\usepackage{algpseudocode}

\ifCLASSINFOpdf

\else
 
\fi

\begin{document}

\title{\LARGE DMRS-Based Uplink Channel Estimation for MU-MIMO Systems with Location-Specific SCSI Acquisition}

\author{Jiawei~Zhuang,
	Hongwei Hou,~\IEEEmembership{Graduate Student Member,~IEEE}, Minjie~Tang, Wenjin~Wang,~\IEEEmembership{Member,~IEEE}, Shi~Jin,~\IEEEmembership{Fellow,~IEEE}, Vincent K. N. Lau,~\IEEEmembership{Fellow,~IEEE}
	\thanks{
		 Jiawei Zhuang, Hongwei Hou, and Wenjin Wang are with the National Mobile Communications Research
		Laboratory, Southeast University, Nanjing 210096, China, and also
		with Purple Mountain Laboratories, Nanjing 211100, China (e-mail:
        \{jw-zhuang, hongweihou,wangwj\}@seu.edu.cn).} 
			\thanks{
			Minjie Tang is with the Department of Communication Systems, EURECOM, France (e-mail: Minjie.Tang@eurecom.fr).} 
					\thanks{
		Shi Jin is with the National Mobile Communications
		Research Laboratory, Southeast University, Nanjing 210096, China (e-mail: jinshi@seu.edu.cn).} 
			\thanks{
		  Vincent K. N. Lau is with the Department of Electronic and Computer
		Engineering, Hong Kong University of Science and Technology, Hong Kong
		(e-mail:  eeknlau@ust.hk).}

}

\maketitle

\begin{abstract}

With the growing number of users in multi-user multiple-input multiple-output (MU-MIMO) systems, demodulation reference signals (DMRSs) are efficiently multiplexed in the code domain via orthogonal cover codes (OCC) to ensure orthogonality and minimize pilot interference.
 In this paper, we investigate uplink DMRS-based channel estimation for MU-MIMO systems with Type II OCC pattern standardized  in third generation partnership project (3GPP) Release 18, leveraging location-specific statistical channel state information (SCSI)  to enhance performance.
Specifically, we propose a SCSI-assisted Bayesian channel estimator (SA-BCE) based on the minimum mean square error  criterion to suppress the pilot interference and noise, albeit at the cost of cubic computational complexity due to matrix inversions. To reduce this complexity while maintaining performance, we extend the scheme to a windowed version (SA-WBCE), which incorporates  antenna-frequency domain windowing and  beam-delay domain processing to exploit asymptotic sparsity and mitigate energy leakage in practical systems.
 To avoid the frequent real-time SCSI acquisition, we construct a grid-based location-specific SCSI database based on the principle of spatial consistency, and subsequently  leverage the uplink received signals within each grid to extract the SCSI. Facilitated by the multilinear structure of wireless channels, we formulate  the SCSI acquisition problem within each grid as a tensor
 decomposition problem, where the factor matrices are parameterized by the multi-path powers, delays, and angles. The computational complexity of SCSI acquisition can be significantly reduced by exploiting the Vandermonde structure of the factor matrices.
 Simulation results demonstrate that the proposed location-specific SCSI database construction method achieves high accuracy, while the SA-BCE and SA-WBCE significantly outperform state-of-the-art benchmarks in MU-MIMO systems.

\end{abstract}
\begin{IEEEkeywords}
MU-MIMO,  SCSI, channel estimation, tensor decomposition, demodulation reference signal
\end{IEEEkeywords}

\IEEEpeerreviewmaketitle

\section{Introduction}
\IEEEPARstart{D}{R}{I}{V}{E}{N}  by the growing demand for data-hungry applications such as broadband internet of things (IoT)\cite{CiteBroadIoT} and extended reality (XR)\cite{CiteXR}, uplink-centric broadband communication (UCBC) \cite{Cite5GAUCBCHuaWei} has emerged as a crucial service category in the era of fifth generation (5G)-Advanced\cite{CiteB5GUCBC1}\cite{CiteB5GUCBC2}. To boost the spectral efficiency of uplink transmission, the multi-user multiple-input multiple-output (MU-MIMO) communication in  UCBC  is required to support a larger number of users  transmitting data concurrently over the same resource element (RE) through spatial multiplexing. Specifically, to meet the communication requirements of more data streams in UCBC, the number of orthogonal pilot ports  increases from $12$ to $24$ in third generation partnership project (3GPP) Release $18$\cite{Cite38.211R18}, supporting up to $24$  users for uplink transmission.

However, in the uplink transmission, the  channel state information (CSI) of multiple users is estimated using the uplink demodulation reference signal (DMRS)\cite{Cite38.211R16},  whose performance degrades as the number of users increases due to the overlap of DMRS from multiple users on the same RE, thereby constraining the spectral efficiency of the MU-MIMO system.
The improvement of  DMRS-based channel estimation is thus indispensable for achieving reliable and efficient uplink MU-MIMO transmissions.

\subsection{Prior Work}

In MU-MIMO uplink transmission, orthogonal cover codes (OCC) are typically employed to enable code-domain multiplexing of DMRS  from multiple users, resulting in their overlap on the same RE. Therefore, a critical step in uplink DMRS channel estimation is  to separate the DMRS  of different users through  OCC decomposition. Based on the assumption that the channels remain unchanged over consecutive subcarriers, many works on uplink DMRS channel estimation leverage the  orthogonality of different pilot ports to achieve OCC decomposition\cite{CiteVSDTensorDecompositionMethod}\cite{CiteMUMIMODMRS}. However, this assumption breaks down in the presence of  frequency-selective fading channel, compromising the orthogonality among different pilot ports and consequently introducing pilot interference.  To address this issue, 
the author in \cite{TwoPortDMRSChannelEstimation} proposed an minimum mean square error (MMSE) scheme for two-port DMRS channel estimation in New Radio systems, which exploits frequency-domain channel correlations to facilitate OCC decomposition.

However, the method in \cite{TwoPortDMRSChannelEstimation} assumes that statistical CSI (SCSI) is perfectly known, which is unrealistic in practical systems. 
 To this end, several SCSI acquisition approaches have been proposed in massive MIMO systems\cite{Citesun2015beamSCSI,Citezheng2015statisticalSCSI,Citevila2013expectationSCSI,Citewen2014channelSCSI,CiteSRSSCSI,Citewang2018statisticalSCSI,Citelu20232dSCI}. In \cite{Citesun2015beamSCSI} and \cite{Citezheng2015statisticalSCSI}, the instantaneous CSI (ICSI) is first estimated through pilot transmission, and SCSI is then derived by performing statistical calculations on the estimated  ICSI.  Expectation-maximization (EM) algorithm \cite{Citevila2013expectationSCSI} \cite{Citewen2014channelSCSI} \cite{CiteSRSSCSI} provides another method for obtaining SCSI, where the ICSI and SCSI are iteratively estimated. Some studies \cite{Citewang2018statisticalSCSI}\cite{Citelu20232dSCI} also proposed methods for acquiring SCSI without the need for  ICSI.  In\cite{Citewang2018statisticalSCSI},  a hidden statistical channel state Markov model (HSCSM model) was proposed for SCSI acquisition in nonstationary massive MIMO environments, where parameters are estimated from received signals and SCSI is obtained via a maximum a-posteriori decision process. The author in \cite{Citelu20232dSCI} proposed a novel approach for obtaining the beam-domain channel power matrices based on the received signal model and Kullback-Leibler divergence.

The aforementioned SCSI acquisition methods primarily rely on real-time received signals for SCSI estimation, which leads to large reference signal overhead and processing delay.
Owing to the intrinsic relationship between  SCSI and the wireless channel’s scattering environment, a consistent mapping  exists between user location and SCSI in stationary environments.
This motivates the construction of location-specific SCSI database \cite{Citezeng2021toward6GCKM}\cite{Citezeng2024CKMtutorialSCSI}, thereby converting the real-time estimation of SCSI into the problem of mapping construction.
Toward this end, various location-specific SCSI database construction methods have been proposed in \cite{CiteZengyonCKMConstruction} and \cite{CiteCKMConstructionI2IYouLi}, which leverage error-free SCSI at specific locations to infer the location-specific SCSI over the entire region via   EM-based  interpolation algorithm and  Laplacian pyramid–based image-to-image inpainting approach, respectively.

Current approaches to building location-specific SCSI databases \cite{CiteZengyonCKMConstruction,CiteCKMConstructionI2IYouLi} often rely on the idealized assumption of error-free SCSI samples, which rarely holds in practical deployment scenarios.
This motivates the investigation of location-specific SCSI database construction methods that exploit noisy received signals at the base station (BS).
 However, the use of  noisy received signals at the BS introduces new challenges. The received signal-to-noise ratio (SNR) at the BS is limited due to the constrained transmitted power of user equipment. In addition, the received signals exhibit a nonlinear relationship with the underlying SCSI parameters, such as the delay response vector\cite{CiteWUchannelPredictionDelayResponseVector}. 
To address these issues, the author in \cite{CiteYuanXiaojunCKM}  employed Bayesian inference to suppress the noise and resolved the nonlinear mapping between received signals and SCSI parameters,   paving the way for location-specific SCSI acquisition using  the noisy BS received signals.

\subsection{Motivation and Main Contributions}
The state-of-the-art primarily exploits the orthogonality of OCC to enable OCC decomposition under the assumption of frequency flat-fading, which becomes invalid in propagation environments with significant delay spread.
Meanwhile, efforts to enhance channel estimation via location-specific SCSI often rely on idealized, error-free channel samples, underscoring the need to incorporate noisy received signals for improved practical relevance.
Moreover, as future communication systems scale in both antenna array size and bandwidth, constructing location-specific SCSI database becomes increasingly complex, necessitating the development of low-complexity solutions.

In this paper, we investigate the SCSI-assisted DMRS-based channel estimation in MU-MIMO systems and develop the corresponding location-specific SCSI database construction algorithm. The main contributions of this work are summarized as follows:

\begin{itemize}
	\item
By incorporating the Type II DMRS configuration
in 3GPP Release $18$, we develop the signal model for uplink  DMRS-based channel estimation. 
The code-domain multiplexing of DMRS from multiple users on the same RE results in pilot interference,  especially in frequency-selective fading channels.
To mitigate such pilot interference, we formulate the uplink multi-user DMRS-based channel estimation problem based on the MMSE  criterion, leveraging the statistical characteristics of the channel to effectively suppress the pilot interference and noise.

	\item  
	Building on the MMSE formulation,
     we propose a SCSI-assisted Bayesian channel estimator (SA-BCE)   to achieve the DMRS-based channel estimation. Specifically, SA-BCE employs a frequency-domain MMSE estimator to mitigate the pilot interference, followed by an antenna-domain MMSE estimator to further suppress the noise. To reduce the computational complexity of SA-BCE, we shift from the antenna-frequency domain to the beam-delay domain and extend the scheme to windowed SA-BCE (SA-WBCE) to alleviate the energy leakage resulting from the finite number of subcarriers and antennas.

	\item  By exploiting the intrinsic relationship between  SCSI and the scattering environment of wireless channel, we transform the SCSI acquisition problem into a mapping construction problem between SCSI and location.
To reduce the storage overhead of location-specific SCSI, the coverage area of the BS is divided into spatial grids, where the locations within each grid share the common SCSI.
By collecting the noisy received signals within each grid, we formulate the mapping construction problem as a tensor decomposition problem, where the factor matrices are parameterized by the multi-path powers, delays, and angles. By exploiting the Vandermonde structure of the factor matrices, the SCSI within each grid can be acquired via Vandermonde-structured tensor decomposition (VSTD) algorithm, with tensor operations significantly reducing the computational complexity.

\end{itemize}

{\it{Organization}}: The remainder of this paper is organized as follows. Section \ref{sec_system} describes the system model. The SA-BCE and SA-WBCE, along with the location-specific SCSI database, are detailed in Section \ref{sec_SCSIBasedCE}. In Section  \ref{sec_TensorBasedSCSIDatabaseConstruction}, we formulate the location-specific SCSI database construction as a tensor decomposition problem and develop a VSTD-based SCSI database construction method.  Simulation results are given in Section \ref{sec_Simulation}. Finally, Section \ref{sec_Conclusion} concludes this paper.

{\it{Notations}}:  $\mathbf{I}_M $ is the $ M \times M$ identity matrix. The imaginary unit is represented by $\bar{\jmath} = \sqrt{-1}$. $x, \mathbf{x},
 \mathbf{X}$, and $\boldsymbol{\mathcal{X}}$ denote scalars, column vectors, matrices, and tensors, respectively. The superscripts $\{\cdot\}^*$, $\{\cdot\}^T$,$\{\cdot\}^H$, $\{\cdot\}^{-1}$ and  $(\cdot)^{\dagger}$ denote the conjugate, transpose, conjugate transpose, inverse and pseudo-inverse respectively.  $||\cdot||_{\mathrm{F}}$ and $||\cdot||_2$ denote the Frobenius norm and the $l_2$ norm, respectively. 
$\otimes$ , $\odot$ and $\circ$   denote the Kronecker, Khatri-Rao and outer products, respectivelv. $\mathcal{C N}\left(\mu, \sigma^2\right)$ denotes a Gaussian distribution with mean $\mu$ and variance $\sigma^2$. The Kronecker delta function is represented by $\delta[\cdot]$.
$\mathbb{E}\{\cdot\}$ denotes the statistic expectation. $\operatorname{diag}\{\cdot\}$ denote the diagonal  operator.
$\mathrm{r}(\mathbf{A})$ and $\mathrm{kr}(\mathbf{A})$ denote the rank and Kruskal-rank of $\mathbf{A}$, respectively.
$[\mathbf{A}]_{:,m}$, $[\mathbf{A}]_{m: n,:}$ and $[\mathbf{A}]_{:, m: n}$ denote  the $m$-th column of $\mathbf{A}$, the submatrix of $\mathbf{A}$ from the $m$-th to the $n$-th rows, and the submatrix of $\mathbf{A}$ from the $m$-th to the $n$-th columns, respectively.   
$[\cdot]_{i_1, \ldots, i_D}$ is the $\left(i_1, \ldots, i_D\right)$-th element of $D$-order tensor.  $\angle$ denotes the operator for extracting the phase angle.

\section{System Model}
\label{sec_system}
We consider a single-cell massive multiple-input multiple-output orthogonal frequency-division multiplexing (MIMO-OFDM)  uplink system, where the BS is equipped with a uniform planar array (UPA) with half-wavelength antenna spacing and serves $K$ users with  an omni-directional antenna. The UPA with $M=M_{\mathrm{v}} M_{\mathrm{h}}$ antennas comprises $M_{\mathrm{v}}$ and $M_{\mathrm{h}}$ antennas in vertical and horizontal directions, respectively. OFDM modulation is employed with \(N_{\mathrm{FFT}}\) subcarriers, and the number of subcarriers for data transmission is \(N_{\mathrm{c}}\), with a subcarrier spacing of \(\Delta f\).
The system sampling interval and OFDM symbol duration are given by \(T_{\mathrm{sam}} = \frac{1}{N_{\mathrm{FFT}} \Delta f}\) and \(T_{\mathrm{sym}} = \frac{1}{\Delta f}\), respectively.

\subsection{Uplink DMRS Configuration}

\begin{figure}[!t]
	\centering    
	
	\subfigure[] 
	{
			\includegraphics[scale=0.7]{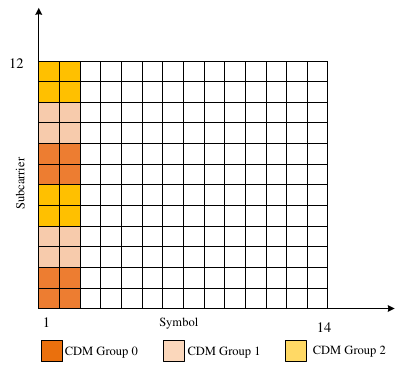} 
	}
	\subfigure[] 
	{
			\includegraphics[scale=0.7]{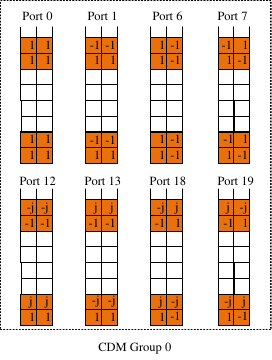}   
	}
	
	\caption{DMRS mapping on OFDM resource grid: (a) Type II DMRS with three CDM groups composed of $24$ orthogonal pilot ports (b) OCC of CDM group $0$ } 
	\label{fig:DMRSIllustration}  
\end{figure}

In the physical uplink shared channel (PUSCH), DMRS is employed to acquire  CSI for subsequent coherent detection. 
 In MU-MIMO systems, the pilot signals transmitted from different users are superimposed at the receiver. To extract  individual signals from the superimposed signal, the DMRS  is scheduled to support multiple orthogonal ports at the transmitter. In 3GPP Release 18, Type II DMRS is specified to support a larger number of users, which divides the resource grid  into multiple code division multiplexing (CDM) groups, wherein OCC are employed within each CDM group to distinguish the different orthogonal pilot ports. In particular, the maximum number of orthogonal pilot ports for  Type II DMRS in  3GPP Release $18$ is $24$, supporting up to $24$  users for uplink transmission.
Fig. \ref{fig:DMRSIllustration} illustrates the pilot pattern of Type II DMRS in a resource block, where the pilot pattern remains consistent across all resource blocks. Within each CDM group, frequency-domain OCC and time-domain OCC are assigned to adjacent OFDM symbols to maintain orthogonality. In this pilot configuration, the number of subcarriers occupied by each CDM group is $ N = \frac{N_{\mathrm{c}}}{G}$, where $ G $ represents the number of CDM groups.  The number of OFDM symbols occupied by the pilots is $ T_{\mathrm{p}}$.

\subsection{Signal Model}

Under the aforementioned pilot configuration, the number of users in each CDM group is $ \frac{K}{G} $. Without loss of generality, we assume that  users  with index of $\{\frac{iK}{G}+1,\frac{i K}{G}+2,...,\frac{(i+1)K}{G}\} $  are assigned to CDM group $i$, where $i=0, 1 ,..., G-1$.
The set of pilot subcarrier indices for CDM group $i$ is denoted as $\mathcal{P}_{i}=\left\{\left.i+1+6\left(n-1\right),i+2+6\left(n-1\right) \right\rvert\, n=1,2 ,\cdots ,\frac{N}{2} \right\}$. 

Consequently, the received signal $ \mathbf{Y}_{i}(t) \in \mathbb{C}^{N \times M} $ of CDM group $ i $ in the $t$-th symbol duration at the BS can be expressed as
\begin{equation}
	\mathbf{Y}_{i}(t) = \sum_{k=\frac{iK}{G}+1}^{\frac{(i+1)K}{G}} w_{t,k}\mathbf{S}_{i}\mathbf{C}_{k} \mathbf{P}_{i}\mathbf{H}_{k}(t) + \mathbf{N}(t),
\end{equation}
where $\mathbf{H}_{k}(t) \in \mathbb{C}^{N_{\mathrm{c}} \times M} $ represents  the frequency-space domain channel  for the $k$-th user at the $t$-th symbol duration, as  detailed in Section  \ref{Channel Model}.  \( \mathbf{P}_{i} \in \{0, 1\}^{N \times N_{\mathrm{c}}} \) is a sampling matrix used to select the pilot subcarrier set \( \mathcal{P}_i \).
$\mathbf{C}_k=\operatorname{diag}\left(1,e^{\bar{\jmath} 2 \pi \frac{\Delta_k}{N}} \ldots, e^{\bar{\jmath} 2 \pi \frac{(N-1)\Delta_k}{N}} \right)\in \mathbb{C}^{N \times N} $ represents the frequency-domain OCC of the \( k \)-th user, where \( \Delta_k \in \{0, \frac{N}{4}, \frac{N}{2}, \frac{3N}{4}\} \) represents the cyclic shift of the \( k \)-th user, employed to ensure the orthogonality of the DMRS across different users.
$ \mathbf{S}_{i} = \operatorname{diag}\{\mathbf{s}_{i}\} \in \mathbb{C}^{N \times N} $ represents the pilot matrix of the CDM group \( i \),  satisfying $ \mathbf{S}_i \mathbf{S}_i^H = \mathbf{I}_{N} $ due to the unit power of the pilot symbols.
$\mathbf{N}(t)\in  \mathbb{C}^{N \times M} $ is the complex Gaussian noise  consisting of independently and identically distributed (i.i.d.) Gaussian variables which follow the i.i.d. complex Gaussian distribution $\mathcal{C N}\left(0, \sigma\right)$. $ w_{t,k} \in \{1, -1\} $ represents the time-domain OCC of the $ k $-th user at the $ t $-th symbol duration, which satisfies\footnote{
	According to the DMRS configuration in 3GPP Release $18$, half of the users within a CDM group share the same time-domain OCC \cite{Cite38.211R18}. Without loss of generality, we assume that the users in each CDM group are divided into two sets based on whether their time-domain OCC are identical: the first set consisting of users indexed from \(\frac{iK}{G}+1\) to \(\frac{(2i+1)K}{2G}\), and the second set consisting of users indexed from \(\frac{(2i+1)K}{2G}+1\) to \(\frac{(i+1)K}{G}\). In the main context, we consider the users in the first set without loss of generality, i.e., $u \in \left\{ \frac{iK}{G}+1, \cdots, \frac{(2i+1)K}{2G} \right\}$.}
\begin{equation}
	\frac{1}{T_p}\sum_{t=1}^{T_p}w_{t,p}w_{t,q} = 
	\begin{cases} 
		1, & \text{if } (p,q \in \mathcal{U}_1) \, \text{or}\, (p,q \in \mathcal{U}_2),  \\
		0, & \text{otherwise},
	\end{cases}
	\label{TimeDomainOCCOrthogonal}
\end{equation} 
where $\mathcal{U}_{1}=\{ \frac{iK}{G}+1, \cdots, \frac{(2i+1)K}{2G}\}$, $\mathcal{U}_{2}=\{\frac{(2i+1)K}{2G}+1, \cdots, \frac{(i+1)K}{G}\}$.

\subsection{Channel Model}
\label{Channel Model}

The space-frequency domain channel of \(k\)-th user at \(t\)-th symbol duration can be modeled by \cite{CiteMultiPathChannelModel}
\begin{equation}
	\begin{aligned}
		\mathbf{H}_{k}(t) &= \sum_{l=1}^{L_{t,k}} \alpha_{l,t,k}  \mathbf{b}_{N_{\mathrm{c}}} \left(\tau_{l,t,k}\right)\mathbf{a}(\varphi_{l,t,k}, \theta_{l,t,k})^{T},
		\label{channel model}
	\end{aligned}
\end{equation}
where  $\mathbf{a}(\varphi_{l,t,k}, \theta_{l,t,k})=\mathbf{a}_{\mathrm{v}}(\theta_{l,t,k}) \otimes \mathbf{a}_{\mathrm{h}}(\varphi_{l,t,k}; \theta_{l,t,k}) \in \mathbb{C}^{ M \times 1}$, and $L_{t,k}$ denotes the number of multi-paths for the $k$-th user at the $t$-th symbol duration. $\alpha_{l,t,k}$, \(\tau_{l,t,k}\), \(\theta_{l,t,k}\), and \(\varphi_{l,t,k}\) represent the complex gain, the path delay, the elevation angle of arrival (AoA), and the azimuth AoA of the \(l\)-th path  for  $k$-th user during the $t$-th symbol duration, respectively.  $\mathbf{b}_{N_{\mathrm{c}}}(\tau) \in \mathbb{C}^{N_{\mathrm{c}} \times 1}$, $\mathbf{a}_{\mathrm{v}}(\theta) \in \mathbb{C}^{M_{\mathrm{v}} \times 1}$ and $\mathbf{a}_{\mathrm{h}}(\varphi ; \theta) \in \mathbb{C}^{M_{\mathrm{h}} \times 1}$ represent the steering vectors in the delay, elevation angle and azimuth angle domains, which are  defined as  $\left[\mathbf{b}_{N_{\mathrm{c}}}(\tau)\right]_n=e^{-\bar{\jmath} 2 \pi n \Delta f \tau}$, $\left[\mathbf{a}_{\mathrm{v}}(\theta)\right]_n= e^{-\bar{\jmath} \pi n \cos \theta}$ and $\left[\mathbf{a}_{\mathrm{h}}(\varphi, \theta)\right]_n= e^{-\bar{\jmath} \pi n \sin \theta \cos \varphi}$, respectively.

We assume an uncorrelated fading environment, where different propagation paths are independent.  Therefore, the complex gain satisfies\cite{CiteChannelYOULIUncorrelatedChannel}
\begin{equation}
	\mathrm{E}\left[\alpha_{l,t,k} \alpha_{l^{\prime},t,k}^*\right]=\rho_{l,t,k}\delta\left[l-l^{\prime}\right],
	\label{UncorrelatedChannel}
\end{equation}
where \(\rho_{l,t,k}\) denotes the power of the \(l\)-th path for the \(k\)-th user during the \(t\)-th symbol duration.

\section{SCSI-Assisted Channel Estimation and Location-Specific SCSI Database }
\label{sec_SCSIBasedCE}
The channel estimation problem  for  CDM group $i$ can be formulated as the  following optimization problem,  given by
\begin{equation}
	\min _{\left\{\mathbf{H}_{k}(t)\right\}_{k=\frac{iK}{G}+1}^{\frac{(i+1)K}{G}}} \sum_{t=1}^{T_p}\left\|\mathbf{Y}_{i}(t)-\sum_{k=\frac{iK}{G}+1}^{\frac{(i+1)K}{G}}w_{t,k}\mathbf{S}_{i}\mathbf{C}_{k} \mathbf{H}^{\text{PS}}_{k}(t) \right\|_{\mathrm{F}}^2,
	\label{CEOptimizeProblem}
\end{equation}
where $ \mathbf{H}^{\text{PS}}_{k}(t)=  \mathbf{P}_{i}\mathbf{H}_{k}(t)\in \mathbb{C}^{N \times M} $ represents the frequency-space domain channel of the pilot segment for the $k$-th user at the $t$-th symbol duration. In the above optimization problem, the number of channel parameters to be estimated is $\frac{KN_{\mathrm{c}}MT_p}{G}$, whereas  the number of received  signal is $NMT_p$,  resulting in an under-determined estimation problem.

The trivial approach \cite{CiteVSDTensorDecompositionMethod} to solving \eqref{CEOptimizeProblem} consists of least-squares (LS) channel estimation, followed by time- and frequency-domain OCC decomposition, and finally, linear interpolation.
However, this approach faces challenges in frequency-selective fading channels and becomes increasingly ineffective as the number of users grows.
   On one hand, it assumes that the channel coefficients remain unchanged over consecutive  subcarriers, which does not hold under frequency-selective fading channels. On the other hand, since the trivial approach does not incorporate inter-subcarrier correlations in  OCC decoupling, its channel estimation performance deteriorates with increased pilot interference.
To address these limitations, we first propose SA-BCE and SA-WBCE  based on the MMSE criterion. Subsequently,  we propose a location-specific SCSI database to facilitate efficient SCSI acquisition.

\subsection{SCSI-Assisted Bayesian Channel Estimator}
\label{SA-BCE}
To address the  limitations of the trivial approach, we propose the SA-BCE, which retains the first and second steps of the trivial method while introducing improvements in the subsequent stages. Specifically, the received signal $\mathbf{Y}_{i}(t)$ is first  divided by the pilot signal $\mathbf{S}_{i}$ to perform LS channel estimation, as shown by
 \begin{equation}
 		\hat{\mathbf{Y}}_{i}^{\text{LS}}(t)=\mathbf{S}_{i}^{H}\mathbf{Y}_{i}(t)=\sum_{k=\frac{iK}{G}+1}^{\frac{(i+1)K}{G}}w_{t,k}\mathbf{C}_{k} \mathbf{H}^{\text{PS}}_{k}(t) +\mathbf{S}_{i}^{H}\mathbf{N}(t),
 \end{equation}
where \(\hat{\mathbf{Y}}_{i}^{\text{LS}}(t) \in \mathbb{C}^{N \times M}\) denotes the LS channel estimation result for the pilot segment during the \(t\)-th symbol duration.

Assuming that the channel coefficients remain constant across consecutive OFDM symbols\footnote{Considering the velocity of the user $v=3$ km/h, the carrier frequency $f_{c}=6.7$ GHz, the coherence time  \cite{CiteMultiPathChannelModel} $T_c$ is around $20$ ms, which is much longer than the OFDM symbol duration $T_\mathrm{sym}=33.3$us for $\Delta f=30$ kHz\cite{Cite38.211R18}. Consequently,  the channel coefficients can be assumed to remain unchanged between consecutive OFDM symbols.}, the effect of the time-domain OCC in $\hat{\mathbf{Y}}_{i}^{\text{LS}}(t)$  can be eliminated by exploiting its orthogonality property, as shown in equation \eqref{TimeDomainOCCOrthogonal}, through
\begin{equation}
	\hat{\mathbf{Y}}_{i}^{\text{LS}}= \frac{1}{T_p}\sum_{t=1}^{T_p}w_{t,u}	\hat{\mathbf{Y}}_{i}^{\text{LS}}(t)=\sum_{k \in \mathcal{U}_{1} }\mathbf{C}_{k} \mathbf{H}^{\text{PS}}_{k}+\mathbf{Z},
\end{equation}
where $\mathbf{Z} = \frac{1}{T_p}\sum_{t=1}^{T_p}w_{t,u}\mathbf{S}_{i}^{H}\mathbf{N}(t) \in \mathbb{C}^{N \times M}$.

 After the time-domain OCC decoupling, in contrast to the trivial approach which  leverages the  orthogonality of  frequency-domain OCC to achieve OCC decomposition, the MMSE criterion is employed to decompose the frequency-domain OCC in $\hat{\mathbf{Y}}_{i}^{\text{LS}}$. The frequency-domain signal\footnote{Since the processing on different receive antennas is the same during the decomposition of the frequency-domain OCC, we omit the antenna index for convenience without loss of generality.}  $\hat{\mathbf{y}}_{i}^{\text{LS}} \in \mathbb{C}^{N \times 1}$  after decomposing the time-domain OCC can be expressed as
\begin{equation}
	\hat{\mathbf{y}}_{i}^{\text{LS}}=\sum_{k=\frac{iK}{G}+1}^{\frac{(2i+1)K}{2G}}\mathbf{C}_{k} \mathbf{h}^{\text{PS}}_{k}+\mathbf{z},
\end{equation}
where $\mathbf{h}^{\text{PS}}_{k} \in \mathbb{C}^{N \times 1}$ represents the frequency-domain channel of the pilot segment for the $k$-th user. The MMSE channel estimation of $\mathbf{h}^{\text{PS}}_u$ can then be obtained  according to \cite{CiteMMSEChannelEstiamtion}:
\begin{equation}
		\hat{\mathbf{h}}_{u}^{\text{PS}}=\mathbf{R}_{\mathbf{h}^{\text{PS}}_{u} \hat{\mathbf{y}}_{i}^{\text{LS}}} \mathbf{R}_{\hat{\mathbf{y}}_{i}^{\text{LS}} \hat{\mathbf{y}}_{i}^{\text{LS}}}^{-1} \hat{\mathbf{y}}_{i}^{\text{LS}},
	\label{frequencyMMSE}
\end{equation}
where $\mathbf{R}_{\mathbf{h}^{\text{PS}}_{u} \hat{\mathbf{y}}_{i}^{\text{LS}}}\in \mathbb{C}^{N \times N}$ and $\mathbf{R}_{\hat{\mathbf{y}}_{i}^{\text{LS}} \hat{\mathbf{y}}_{i}^{\text{LS}}}\in \mathbb{C}^{N \times N}$ can be expressed as
\begin{equation}
	\begin{aligned}
		\mathbf{R}_{\mathbf{h}^{\text{PS}}_{u} \hat{\mathbf{y}}_{i}^{\text{LS}}}&=\mathbb{E}\left\{\mathbf{h}^{\text{PS}}_{u} \left(\hat{\mathbf{y}}_{i}^{\text {LS }}\right)^H\right\}=\mathbf{P}_{i}\mathbf{R}^{f}_{u}\mathbf{P}_{i}^{T}\mathbf{C}_u^H,\\
		\mathbf{R}_{\hat{\mathbf{y}}_{i}^{\text{LS}} \hat{\mathbf{y}}_{i}^{\text{LS}}}&=\mathbb{E}\left\{\hat{\mathbf{y}}_{i}^{\text {LS }}\left(\hat{\mathbf{y}}_{i}^{\text {LS }}\right)^H\right\}\\
		&=\sum_{k=\frac{iK}{G}+1}^{\frac{(2i+1)K}{2G}} \mathbf{C}_k \mathbf{P}_{i}\mathbf{R}^f_{k} \mathbf{P}_{i}^{T}\mathbf{C}_k^H+\sigma^2 \mathbf{I}_{N}.
		\label{correlationdefinition}
	\end{aligned}
\end{equation}
Here, $\mathbf{R}_{k}^{f} \in \mathbb{C}^{N \times N}$ denotes  the frequency-domain channel correlations  for the $k$-th user and is expressed as
\begin{equation}
	\mathbf{R}^{f}_{k}=\sum_{l=1}^{L_{t,k}} \rho_{l,t,k} \mathbf{b}_{N_{\mathrm{c}}} \left(\tau_{l,t,k}\right) \mathbf{b}_{N_{\mathrm{c}}}^H\left(\tau_{l,t,k}\right).
	\label{frequency-domain channel correlation matrix definition}
\end{equation}

As a result,  $\{\hat{\mathbf{H}}^{\text{PS}}_{u}  \}_{u=1}^{K}$ can then be obtained by applying \eqref{frequencyMMSE} to the received signals $\hat{\mathbf{y}}_{i}^{\text{LS}}$ from all CDM groups and receive antennas. To further suppress the noise in $\hat{\mathbf{H}}^{\text{PS}}_{u}$, we exploit the inter-antenna correlations and employ the MMSE channel estimation in the antenna domain, which gives \cite{CiteMMSEChannelEstiamtion}
\begin{equation}
	 (\widetilde{\mathbf{H}}^{\text{PS}}_{u})^{T} =\mathbf{R}^{s}_{u}(\mathbf{R}^{s}_{u}+\sigma^2\mathbf{I}_{M})^{-1}(\hat{\mathbf{H}}^{\text{PS}}_{u})^{T},
	\label{FinalantennaMMSE}
\end{equation}
where $\mathbf{R}^{s}_{u} \in \mathbb{C}^{M \times }$ is the antenna domain channel correlations for the $u$-th user, expressed as
\begin{equation}
	\begin{aligned}
		\mathbf{R}^{s}_{u}=\sum_{l=1}^{L_{t,u}} \rho_{l,t,u}  \mathbf{a}(\varphi_{l,t,u}, \theta_{l,t,u})\mathbf{a}^{H}(\varphi_{l,t,u}, \theta_{l,t,u}).
	\end{aligned}
	\label{antenna-domain channel correlation matrix definition}
\end{equation}

Finally,  the full-frequency domain channel for all users, represented as \( \{\hat{\mathbf{H}}^{\mathrm{MMSE}}_{u} \in \mathbb{C}^{N_{\mathrm{c}} \times M}\}_{u=1}^{K} \), is reconstructed by applying frequency-domain linear interpolation to the pilot segment channel \( \{\widetilde{\mathbf{H}}^{\text{PS}}_{u} \in \mathbb{C}^{N \times M}\}_{u=1}^{K} \).

\subsection{Low-Complexity SA-BCE}
\label{SA-WBCE}
The overall complexity of SA-BCE is dominated by matrix inversions with complexity of $\mathcal{O}(N^3+M^3)$. To reduce the computational complexity of SA-BCE, we implement the SA-BCE in  \eqref{frequencyMMSE} - \eqref{antenna-domain channel correlation matrix definition} in the beam-delay domain \cite{CiteChannelSparsity1}\cite{CiteChannelSparsity2}\cite{BayesianDelayPrecoder}, thereby exploiting the limited scattering nature of wireless channels to reduce complexity. Specifically, the beam-delay domain MMSE estimator is given by
\begin{equation}
	\begin{aligned}
		&\hat{\mathbf{H}}_{u}^{\text{PS}}=\mathbf{C}_u^{H}\mathbf{F}_N\mathbf{R}^{\tau}_{u} (\sum_{k=\frac{iK}{G}+1}^{\frac{(2i+1)K}{2G}}\mathbf{R}^{\tau}_{k}+\sigma^2 \mathbf{I}_{N})^{-1} \mathbf{F}_N^{H}	\hat{\mathbf{Y}}_{i}^{\text{LS}},\\
		& (\widetilde{\mathbf{H}}^{\text{PS}}_{u})^{T}=\mathbf{F}^{A} \mathbf{R}^{a}_{u}(\mathbf{R}^{a}_{u}+\sigma^2 \mathbf{I}_{M})^{-1}\left(\mathbf{F}^{A} \right)^H(\hat{\mathbf{H}}^{\text{PS}}_{u})^{T},\\
	\end{aligned}
	\label{NoWindowBeamDelayDomainEstimator}
\end{equation} 
where $\mathbf{F}_N \in \mathbb{C}^{N \times N}$ denotes the discrete fourier transform (DFT) matrix,  with its $(i,j)$-th entry defined as $\left[\mathbf{F}_N\right]{i, j} \triangleq \frac{1}{\sqrt{N}} e^{-\bar{\jmath} 2 \pi \frac{i j}{N}}$, $\mathbf{F}^{A}=\left(\mathbf{F}_{M_{\mathrm{v}}} \otimes \mathbf{F}_{M_{\mathrm{h}}}\right)$. $\mathbf{R}^{\tau}_{u}\in \mathbb{C}^{N \times N}$ and $\mathbf{R}^{a}_{u}\in \mathbb{C}^{M \times M}$ denote the delay- and beam-domain channel correlations of the $u$-th user, respectively, and are defined as
	\begin{equation}
	\small
	\begin{aligned}
		\mathbf{R}^{\tau}_{u}&=\mathbf{F}_N^H\mathbf{C}_u \mathbf{P}_{i}\mathbf{R}^{f}_{u} \mathbf{P}_{i}^{T}\mathbf{C}_u^H\mathbf{F}_N = \sum_{l=1}^{L_{t,u}} \rho_{l,t,u}\overline{\mathbf{b}}(\tau_{l,t,u})\overline{\mathbf{b}}^{H}(\tau_{l,t,u}), \\
		\mathbf{R}^{a}_{u}&=(\mathbf{F}^{A})^H\mathbf{R}^{s}_{u}\mathbf{F}^{A}= \sum_{l=1}^{L_{t,u}} \rho_{l,t,u}  \overline{\mathbf{a}}(\varphi_{l,t,u}, \theta_{l,t,u})\overline{\mathbf{a}}^{H}(\varphi_{l,t,u}, \theta_{l,t,u}),
	\end{aligned}
	\label{BeamDelayChannelCorrelations}
\end{equation} 
where $\overline{\mathbf{b}}(\tau_{l,t,u})=\mathbf{F}_N^H\mathbf{C}_u \mathbf{P}_{i} \mathbf{b}_{N_{\mathrm{c}}} \left(\tau_{l,t,u}\right)$, $\overline{\mathbf{a}}(\varphi_{l,t,u},\theta_{l,t,u})=\overline{\mathbf{a}}_{\mathrm{v}}(\theta_{l,t,u})\otimes \overline{\mathbf{a}}_{\mathrm{h}}(\varphi_{l,t,u}; \theta_{l,t,u})$, $\overline{\mathbf{a}}_{\mathrm{v}}(\theta_{l,t,u})=\mathbf{F}^{H}_{M_{\mathrm{v}}}\mathbf{a}_{\mathrm{v}}(\theta_{l,t,u})$ and $\overline{\mathbf{a}}_{\mathrm{h}}(\varphi_{l,t,u}; \theta_{l,t,u})=\mathbf{F}^{H}_{M_{\mathrm{h}}}\mathbf{a}_{\mathrm{h}}(\varphi_{l,t,u}; \theta_{l,t,u})$. 	

 Observing that the beam–delay domain MMSE estimator is equivalent to its antenna–frequency domain counterpart, we present the following lemma to establish their equivalence.

\textit{Lemma 1: }The beam-delay domain MMSE estimator is equivalent to its antenna-frequency domain counterpart.

\textit{Proof:} Utilizing the orthogonality property of the DFT matrix  and the relation in \eqref{BeamDelayChannelCorrelations}, substitution of \eqref{frequencyMMSE} and \eqref{FinalantennaMMSE} directly yields the estimator in \eqref{NoWindowBeamDelayDomainEstimator}, thereby proving the equivalence. $\hfill \blacksquare$

Owing to the intrinsic sparsity of the beam–delay domain channel, the matrices $\mathbf{R}^{\tau}_{u}$ and $\mathbf{R}^{a}_{u}$ exhibit sparse structures, thereby enabling a significant reduction in the computational complexity. Specifically, we present the following proposition to characterize the asymptotic behavior of   $\mathbf{R}^{\tau}_{u}$ and  $\mathbf{R}^{a}_{u}$.

\textit{Proposition 1: }In the infinite case, \(\mathbf{R}^{\tau}_{u}\) and \(\mathbf{R}^{a}_{u}\) respectively converge to diagonal matrices.

\textit{Proof:} In the infinite case where both the number of antennas and subcarriers tend to infinite, the sampled steering vectors exhibit asymptotic orthogonality \cite{CiteInfiniteCase}, rendering $\overline{\mathbf{b}}(\tau_{l,t,u})$ and $\overline{\mathbf{a}}(\varphi_{l,t,u},\theta_{l,t,u})$ asymptotically 1-sparse\cite{CiteSRSSCSI}, i.e., containing only a single non-zero element. Consequently, according to \eqref{BeamDelayChannelCorrelations},  both $\mathbf{R}^{\tau}_{u}$ and $\mathbf{R}^{a}_{u}$ reduce to diagonal matrices in this case. This completes the proof. $\hfill \blacksquare$

Proposition 1 reveals that the beam-delay domain MMSE estimator degenerates into a element-wise operation as the number of antennas and subcarriers tends to infinity, thereby eliminating the need for matrix inversions. However, in practical systems, the number of antennas and subcarriers is finite, leading to inevitable energy leakage in the beam-delay domain. Therefore, both the beam- and delay-domain channel correlation matrices exhibit significant off-diagonal elements, which in turn substantially increases the computational complexity of the beam-delay domain MMSE estimator.

To alleviate the energy leakage problem, we incorporate the energy-concentrating property of window functions \cite{CiteWindowIGA}\cite{CiteWindowHFWaveReceiver} into the beam-delay domain MMSE estimator to develop the SA-WBCE,   given by
	\begin{equation}
	\begin{aligned}
		&\hat{\mathbf{H}}_{u}^{\text{PS}}=\boldsymbol{\Lambda}^{-1}_{f}\mathbf{C}_u^{H}\mathbf{F}_N\widetilde{\mathbf{R}}^{\tau}_{u} (\sum_{k=\frac{iK}{G}+1}^{\frac{(2i+1)K}{2G}}\widetilde{\mathbf{R}}^{\tau}_{k}+\sigma^2\boldsymbol{\Xi}_f )^{-1} \mathbf{F}_N^{H}	\boldsymbol{\Lambda}_{f} \hat{\mathbf{Y}}_{i}^{\text{LS}},\\
		&(\widetilde{\mathbf{H}}^{\text{PS}}_{u})^{T}=\boldsymbol{\Lambda}^{-1}_{s}\mathbf{F}^{A} \widetilde{\mathbf{R}}^{a}_{u}(\widetilde{\mathbf{R}}^{a}_{u}+\sigma^2\boldsymbol{\Xi}_s)^{-1}\left(\mathbf{F}^{A} \right)^H\boldsymbol{\Lambda}_{s}(\hat{\mathbf{H}}^{\text{PS}}_{u})^{T},
	\end{aligned}
\end{equation} 
where $\boldsymbol{\Xi}_f=\mathbf{F}_N^{H}|\boldsymbol{\Lambda}_{f}|^2\mathbf{F}_N$, $\boldsymbol{\Xi}_s=\left(\mathbf{F}^{A} \right)^H|\boldsymbol{\Lambda}_{s}|^2 \mathbf{F}^{A}$, $\boldsymbol{\Lambda}_{f} \triangleq \operatorname{diag}\{\boldsymbol{\eta}_{f}\} \in \mathbb{C}^{N \times N}$  and $\boldsymbol{\Lambda}_{s} \triangleq \operatorname{diag}\{\boldsymbol{\eta}_{s}\} \in \mathbb{C}^{M \times M}$ denote the frequency- and antenna-domain window functions, respectively.  \(\widetilde{\mathbf{R}}^{\tau}_{u}\in \mathbb{C}^{N \times N}\) and \(\widetilde{\mathbf{R}}^{a}_{u}\in \mathbb{C}^{M \times M}\) denote the windowed versions of the delay- and beam-domain channel correlations for the \(u\)-th user, respectively, and are given by
\begin{equation}
	\begin{aligned}
		\widetilde{\mathbf{R}}^{\tau}_{u}&=\mathbf{F}_N^H\boldsymbol{\Lambda}_{f}\mathbf{C}_u \mathbf{P}_{i}\mathbf{R}^{f}_{u} \mathbf{P}_{i}^{T}\mathbf{C}_u^H\boldsymbol{\Lambda}^{H}_{f}\mathbf{F}_N\\
		& = \sum_{l=1}^{L_{t,u}} \rho_{l,t,u}\widetilde{\mathbf{b}}(\tau_{l,t,u})\widetilde{\mathbf{b}}^{H}(\tau_{l,t,u}), \\
		\widetilde{\mathbf{R}}^{a}_{u}&=(\mathbf{F}^{A})^H\boldsymbol{\Lambda}_{s}\mathbf{R}^{s}_{u}\boldsymbol{\Lambda}_{s}^{H}\mathbf{F}^{A}\\
		&= \sum_{l=1}^{L_{t,u}} \rho_{l,t,u}  \widetilde{\mathbf{a}}(\varphi_{l,t,u}, \theta_{l,t,u})\widetilde{\mathbf{a}}^{H}(\varphi_{l,t,u}, \theta_{l,t,u}),
	\end{aligned}
	\label{WindowChannelCorrleation}
\end{equation} 
where \(\widetilde{\mathbf{b}}(\tau_{l,t,u})\) and \(\widetilde{\mathbf{a}}(\varphi_{l,t,u}, \theta_{l,t,u})\) represent the windowed versions of \(\overline{\mathbf{b}}(\tau_{l,t,u})\) and \(\overline{\mathbf{a}}(\varphi_{l,t,u}, \theta_{l,t,u})\), respectively, exhibiting improved energy concentration. This implies that  \(\widetilde{\mathbf{R}}^{\tau}_{u}\) and \(\widetilde{\mathbf{R}}^{a}_{u}\) can be approximated as band matrices \(\widehat{\mathbf{R}}^{\tau}_{u}\) and \(\widehat{\mathbf{R}}^{a}_{u}\), i.e.,
\begin{equation}
	\left[	\widehat{\mathbf{R}}^{\phi}_{u}\right]_{i,j} = 
	\begin{cases} 
		\left[	\widetilde{\mathbf{R}}^{\phi}_{u}\right]_{i,j}, & \text{if } |i-j| \leq B_{\phi}, \\
		0, & \text{otherwise},
	\end{cases}
	\label{WindowChannelCorrleationSetZero}
\end{equation} 
where $\phi \in\{\tau, a\}$. Due to the characteristics of the window functions, both \(\boldsymbol{\Xi}_f\) and \(\boldsymbol{\Xi}_s \) are band matrices with narrow band sizes. As a result, the matrix inversions in the SA-WBCE can be simplified to the inversion of band matrices, reducing the computational complexity from $\mathcal{O}(N^3+M^3)$ to \(\mathcal{O}(N B_f^2+M B_s^2)\). The proposed SA-WBCE for $\mathcal{U}_{1}$ is summarized in Algorithm \ref{SCSIAssistedChannelEstimation}.  For $\mathcal{U}_{2}$, the SA-WBCE can be readily obtained by modifying the user index accordingly.

Note that both SA-BCE and  SA-WBCE rely on SCSI $\{{\tau}_{l,t,u}, {\varphi}_{l,t,u}, {\theta}_{l,t,u}, {\rho}_{l,t,u}\}_{l=1}^{L_{t,u}}$. However, in practical systems, the ideal SCSI $\{{\tau}_{l,t,u}, {\varphi}_{l,t,u}, {\theta}_{l,t,u}, {\rho}_{l,t,u}\}_{l=1}^{L_{t,u}}$  is not known a priori and is typically obtained by estimation \cite{Citelu20232dSCI}, leading to significant processing delay due to frequent estimation. Leveraging  the assumptions of spatial consistency and environmental stationarity, we can establish a location-specific SCSI database to facilitate SCSI acquisition, which will be introduced in the following subsection. 

\begin{algorithm}[!t]
	\caption{ SA-WBCE for $\mathcal{U}_{1}$} 
	\label{SCSIAssistedChannelEstimation}
	\hspace*{0in} {\bf {Input:}}  \parbox[t]{\dimexpr\linewidth-\algorithmicindent}{$\{ \{{\tau}_{l,t,u}, {\varphi}_{l,t,u}, {\theta}_{l,t,u}, {\rho}_{l,t,u}\}_{l=1}^{L_{t,u}},\mathbf{C}_{u},w_{t,u}\}_{u=1,\cdots,K}$,\\$\quad\{\mathbf{Y}_{i}(t), \mathbf{S}_{i},\mathbf{P}_{i}\}_{i=0,\cdots,G-1}$, $\sigma^2$, $\boldsymbol{\Lambda}_{f}$, $\boldsymbol{\Lambda}_{s}$, $\boldsymbol{\Xi}_f$, $\boldsymbol{\Xi}_s$}
	
	\begin{algorithmic}[1]
		\For{$u=1,...,K$}
		\State
		Compute $\mathbf{R}^{f}_{u}$ and $\mathbf{R}^{s}_{u}$ via \eqref{frequency-domain channel correlation matrix definition} and \eqref{antenna-domain channel correlation matrix definition};
			\State
			Compute $\widehat{\mathbf{R}}^{\tau}_{u}$ and $\widehat{\mathbf{R}}^{a}_{u}$ using \eqref{WindowChannelCorrleation} and \eqref{WindowChannelCorrleationSetZero};
			   \EndFor	
			   
		\For{$i=0,...,G-1$}
		
		\State
	     $\hat{\mathbf{Y}}^{\text{LS}}_{i}(t)=\mathbf{S}_{i}^{H}\mathbf{Y}_{i}(t)$, $t=1,...,T_{\mathrm{p}}$;
	      \State
	     $\mathbf{\Gamma}=(\sum_{k=\frac{iK}{G}+1}^{\frac{(2i+1)K}{2G}}\widehat{\mathbf{R}}^{\tau}_{k}+\sigma^2 \boldsymbol{\Xi}_f )^{-1}$;

		\For{$u=\frac{iK}{G}+1,...,\frac{(2i+1)K}{2G}$}
		
		\quad \% Time-domain OCC decomposition
		
		\State
	 $\hat{\mathbf{Y}}^{\text{LS}}_{i}= \frac{1}{T_p}\sum_{t=1}^{T_p}w_{t,u}	\hat{\mathbf{Y}}_{i}^{\text{LS}}(t)$;
	 
	 	\quad \parbox[t]{\dimexpr\linewidth-\algorithmicindent}{\% Frequency-domain OCC decomposition}
	 	
		\State
	$
	\hat{\mathbf{H}}^{\text{PS}}_{u}=\boldsymbol{\Lambda}^{-1}_{f}\mathbf{C}_u^{H}\mathbf{F}_N\widehat{\mathbf{R}}^{\tau}_{u} \boldsymbol{\Gamma}	\boldsymbol{\Lambda}_{f}\hat{\mathbf{Y}}^{\text{LS}}_{i}$;
		
		\quad \% Antenna-domain MMSE
		\State
		$\boldsymbol{\Omega}=(\widehat{\mathbf{R}}^{a}_{u}+\sigma^2\boldsymbol{\Xi}_s)^{-1}\left(\mathbf{F}^{A} \right)^H\boldsymbol{\Lambda}_{s}$;
		\State
		$(\widetilde{\mathbf{H}}^{\text{PS}}_{u})^{T}=\boldsymbol{\Lambda}_{s}^{-1}\mathbf{F}^{A} \widehat{\mathbf{R}}^{a}_{u}\boldsymbol{\Omega}(\hat{\mathbf{H}}^{\text{PS}}_{u})^{T}$;
		\State
		Perform  linear interpolation to obtain $\hat{\mathbf{H}}^{\mathrm{MMSE}}_{u}$;
	   \EndFor	
		\EndFor	
	\end{algorithmic}
	\hspace*{0in} {\bf {Output:}}$\{\hat{\mathbf{H}}^{\mathrm{MMSE}}_{u} \in \mathbb{C}^{N{\mathrm{c}} \times M}, u\in\mathcal{U}_{1}\}$.
\end{algorithm}

\subsection{ Location-Specific SCSI Database}
\label{sec_LocationBasedSCSIDatabase}
Advancements in localization systems, such as global positioning system (GPS), laser-based systems, inertial measurement units, and integrated sensing and communication (ISAC) \cite{Citeliu2022ISAC}, enable efficient user localization for communication systems. This progress facilitates the availability of abundant high-quality location-tagged channel data. In this context, we propose a location-specific SCSI database that establishes a mapping between user location and SCSI to support the proposed SA-BCE and SA-WBCE.

Since the BS's position is generally fixed after deployment, the SCSI depends on the user's location at $t$-th OFDM symbol duration (denoted as $q(t)$) and the local propagation environment. With the assumption  that the local propagation environment remains quasi-static over a longer time period $T$ relative to the signal transmission period, i.e., the quasi-static environment assumption,  the SCSI  depends only on the location \( q(t) \). Therefore, we can establish a mapping relationship between the user's  location and the SCSI, i.e.,
\begin{equation}
	\mathcal{C}(\cdot): q(t) \rightarrow\left\{\bar{\tau}_{q(t), l}, \bar{\theta}_{q(t), l}, \bar{\varphi}_{q(t), l}, \bar{\rho}_{q(t), l}\right\}_{l=1}^{L_{q(t)}} \quad t \in \mathbb{T},
\end{equation}
where $\mathbb{T}=\{1, \ldots, T\}$ denotes the set of the OFDM symbol duration. Since the SCSI $\{\bar{\tau}_{q(t), l}, \bar{\theta}_{q(t), l}, \bar{\varphi}_{q(t), l}, \bar{\rho}_{q(t), l}\}_{l=1}^{L_{q(t)}}$  remains unchanged over the duration of \( T \) OFDM symbols, we can construct the SCSI database at the \( t' \)-th (\( t'=1 \)) OFDM symbol duration and use the SCSI to assist channel estimation over the subsequent \( T \) OFDM symbol durations.

However, directly mapping all possible user locations within the physical region to the SCSI incurs significant storage overhead. To address this issue, we divide the coverage area of the BS into multiple grids, where users within the same grid share a common SCSI.  The rationale behind grid partitioning is the spatial consistency of wireless channels which refers to the fact that neighboring spatial positions tend to share similar clusters, leading to spatial correlation in SCSI\cite{CiteSpatialConsistency}. The correlation of SCSI at two spatial positions depends on both the distance between them and the correlation distance of the environments. 
When the distance between the two  positions exceeds the correlation distance, the corresponding SCSI of the two positions become statistically independent, and vice versa can be considered identical. 
Thus,  as long as the grid size is much smaller than the correlation distance, the SCSI at different positions within the same grid can be approximated as identical.

\begin{figure}[!t]
	\centering
	\includegraphics[width=0.48\textwidth]{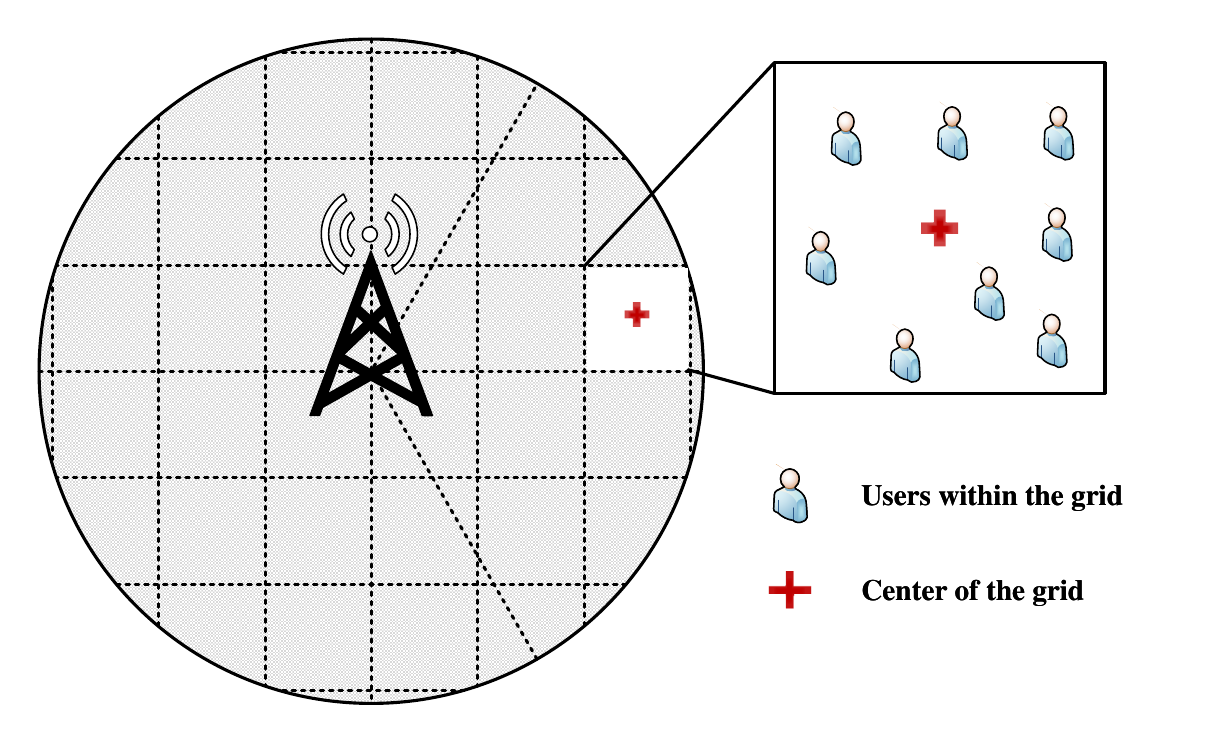}
	\caption{Illustration of the grid-based location-specific SCSI database.}
	\label{fig:GridBasedChannelKnowMap}
\end{figure}

In the following, we provide a detailed description of the grid-based SCSI database as shown in Fig. \ref{fig:GridBasedChannelKnowMap}. Assume that the coverage area of the BS is divided into $ U $ grids, each with a size of $ d \times d \, \text{m}^2 $, where $d$ is much smaller than the correlation distance. Define a mapping $\mathcal{G}(\cdot)$ from the user's location at $t$-th OFDM symbol duration $q(t)$ to a grid $g$ as
\begin{equation}
\label{Map between location and Grid}
\mathcal{L}(\cdot):  g=\mathcal{L}(q(t)) \quad t \in \mathbb{T},g \in \mathbb{G},
\end{equation}
where  $\mathbb{G}=\{1, \ldots, U\}$ denotes the set of the  grid. Based on grid partitioning, we can replace the location-based SCSI $ \{\bar{\tau}_{q(t), l}, \bar{\theta}_{q(t), l}, \bar{\varphi}_{q(t), l}, \bar{\rho}_{q(t), l}\}_{l=1}^{L_{q(t)}} $ with the grid-based one $ \{\bar{\tau}_{g, l}, \bar{\theta}_{g, l}, \bar{\varphi}_{g, l}, \bar{\rho}_{g, l}\}_{l=1}^{\bar{L}} $, where \( \bar{L} \) denotes the number of effective paths of the channel. The corresponding grid-based SCSI database can be represented as
\begin{equation}
	\mathcal{C}(\cdot): g \rightarrow\{\bar{\tau}_{g, l}, \bar{\theta}_{g, l}, \bar{\varphi}_{g, l}, \bar{\rho}_{g, l}\}_{l=1}^{\bar{L}}.
	\label{CKM}
\end{equation}

 To obtain the SCSI of user \( k \)  for  DMRS channel estimation, the position \( q_k(t) \) of user \( k \) at the \( t \)-th OFDM symbol is mapped to the grid \( g_k \) using the function \( g_k = \mathcal{L}(q_k(t)) \) in \eqref{Map between location and Grid}. Subsequently, the SCSI for grid \( g_k \), denoted as \( \{{\tau}_{g_k,l}, {\varphi}_{g_k,l}, {\theta}_{g_k,l}, {\rho}_{g_k,l}\}_{l=1}^{\bar{L}} \), is obtained by applying \( \mathcal{C}(g_k) \) as given in \eqref{CKM}. 
 Based on the obtained SCSI  \( \{{\tau}_{g_k,l}, {\varphi}_{g_k,l}, {\theta}_{g_k,l}, {\rho}_{g_k,l}\}_{l=1}^{\bar{L}} \), the frequency-domain channel correlation matrix $\tilde{\mathbf{R}}^{f}_{k}$ and antenna-domain channel correlation matrix $\tilde{\mathbf{R}}^{s}_{k}$ can be computed using \eqref{frequency-domain channel correlation matrix definition} and \eqref{antenna-domain channel correlation matrix definition}.

Based on \( \tilde{\mathbf{R}}^{f}_{k} \) and \( \tilde{\mathbf{R}}^{s}_{k} \) for all users, the channel can be estimated using the SA-BCE and SA-WBCE as detailed in Section \ref{SA-BCE} and Section \ref{SA-WBCE}, respectively.

\section{VSTD-Based SCSI Database Construction}
\label{sec_TensorBasedSCSIDatabaseConstruction}
In this section, we first analyze the received signals for the location-specific SCSI database construction. Building on this foundation, we reformulate the  location-specific SCSI database  construction as a tensor decomposition problem, where the multilinear structure of wireless channels \cite{CiteMultilinearHouTensor,CiteMultilinearWangTensor} enables a significant reduction in computational complexity.

\subsection{Received Signals for SCSI Database Construction}
The grid-based SCSI database, as shown in Fig. \ref{fig:GridBasedChannelKnowMap}, is constructed by utilizing the received signals from all users within the grid.
 For simplicity, we define \( \mathbb{Q}_g = \{(q, t) \mid g = \mathcal{L}(q), t \in \mathbb{T} \} \) as the set of locations corresponding to grid \( g \) and their corresponding OFDM symbol duration indices. For simplicity, we define \( v = (q,t) \in \mathbb{Q}_g \), as the spatial sampling point for grid \( g \) (the user at location \( q \) during the \( t \)-th OFDM symbol duration). Based on \eqref{CKM}, the channel of the sampling point $v$ located within grid \( g \) can be represented as
\begin{equation}
	\begin{aligned}
		\mathbf{H}_{v} &= \sum_{l=1}^{\bar{L}} \alpha_{l,v} \mathbf{b}_{N_{\mathrm{d}}}\left(\bar{\tau}_{g, l}\right)\mathbf{a}( \bar{\varphi}_{g, l},  \bar{\theta}_{g, l})^{T}+\Delta_{\mathbf{H}_{v}}, v\in \mathbb{Q}_g,
	\end{aligned}
\end{equation}
where \( \mathbf{H}_{v} \in \mathbb{C}^{ N_{\mathrm{d}} \times M} \), and \( N_{\mathrm{d}} \) denotes the number of subcarriers used to obtain the SCSI. $\alpha_{l,v} $ is the \( l \)-th effective complex coefficient, which satisfies $ \mathrm{E}\left[\alpha_{l, v} \alpha_{l, v}^*\right] = \bar{\rho}_{g, l} $. $ \Delta_{\mathbf{H}_{v}} $ represents the error in channel representation and can also indicate the accuracy of the SCSI.  Over the \( T \) OFDM symbol durations, the received signals \( \mathbf{Y}_{v}\in \mathbb{C}^{ N_{\mathrm{d}} \times M} \), \( v \in \mathbb{Q}_g \), are collected as measurement data including sounding reference signals, synchronization signals, etc., from all users within grid \( g \). Subsequently, \( \mathbf{Y}_{v}\), for \( v  \in \mathbb{Q}_g \), is used to construct the SCSI database for grid \( g \), given by
\begin{equation}
	\mathbf{Y}_{v}=\mathbf{S}_{v}	\mathbf{H}_{v} + \tilde{\mathbf{N}}_{v},
\end{equation}
where \( \mathbf{S}_{v}\in \mathbb{C}^{ N_{\mathrm{d}} \times N_{\mathrm{d}}} \) denotes the pilot sequence of the sampling point $v$, $\tilde{\mathbf{N}}_{v}\in \mathbb{C}^{ N_{\mathrm{d}} \times M}$ represents the complex Gaussian noise matrix, where each element follows the i.i.d. complex Gaussian distribution, incorporating the channel representation error induced by $\Delta_{\mathbf{H}_{v}}$.

 To facilitate implementation, we equivalently represent  $\mathbf{Y}_{v}$ in vector form as follows:
\begin{equation}
	\mathbf{y}_{v}=\sum_{l=1}^{\bar{L}} \alpha_{l,v}\left(\mathbf{S}_{v} \mathbf{b}_{N_{\mathrm{d}}}\left(\bar{\tau}_{g, l}\right)\right) \otimes \mathbf{a}( \bar{\varphi}_{g, l},  \bar{\theta}_{g, l}) + \tilde{\mathbf{n}}_{v}.
\end{equation}

Assume that there are \( W \) spatial sampling points \( \{v_1, v_2, \dots, v_W\} \) within each grid, we can use the received signals from these spatial sampling points \( \{\mathbf{y}_{v_1}, \mathbf{y}_{v_2}, \dots,\mathbf{y}_{v_w},\dots, \mathbf{y}_{v_W}, v_{w} \in \mathbb{Q}_g\}_{w = 1}^{W}  \) to obtain the SCSI for  grid $g$. To eliminate the effects of pilot,  we first perform LS channel estimation on the received signal \( \mathbf{y}_{v_w} \), as follows.
\begin{equation}
	\mathbf{h}^{\mathrm{LS}}_{v_w}=	(\mathbf{S}^{H}_{v_w} \otimes \mathbf{I}_{M})\mathbf{y}_{v_w}.
\end{equation}

Let $\mathbf{H}_{g}\in ^{N_{\mathrm{d}}M\times W}$ be the concatenation of the channels from all sampling points, given by
\begin{equation}
	\mathbf{H}_{g}=\left[\mathbf{h}^{\mathrm{LS}}_{v_1},\mathbf{h}^{\mathrm{LS}}_{v_2},\dots,\mathbf{h}^{\mathrm{LS}}_{v_w},\dots,\mathbf{h}^{\mathrm{LS}}_{v_W}\right], v_w\in \mathbb{Q}_g.
\end{equation}
Since we assume that the SCSI remains consistent for each sampling point within each grid, \( \mathbf{H}_g \) can be represented as
\begin{equation}
	\mathbf{H}_{g}=\sum_{l=1}^{\bar{L}}  \left(\mathbf{b}_{N_{\mathrm{d}}}\left(\bar{\tau}_{g, l}\right) \otimes \mathbf{a}( \bar{\varphi}_{g, l},  \bar{\theta}_{g, l})\right)\mathbf{p}^T_{g,l} + \mathbf{N}_{g},
	\label{HgMultipath}
\end{equation}
where \( \mathbf{p}_{g,l} = \left[\alpha_{l,v_1}, \alpha_{l,v_2}, \dots, \alpha_{l,v_W}\right]^{T} \) denotes the channel gains of the \( l \)-th path at all sampling points within grid \( g \). Since \( \alpha_{l,v_w} \) satisfies the condition \( \mathbb{E}[\alpha_{l,v_w} \alpha_{l,v_w}^*] = \bar{\rho}_{l,g} , w = 1, 2, \dots, W\), we can approximate \( \bar{\rho}_{l,g} \) using \( \alpha_{l,v_w} \) as follows:
\begin{equation}
	\bar{\rho}_{l,g}=\frac{1}{W}\sum^{W}_{w=1}|\alpha_{l,v_w}|^2.
\end{equation}

Since the received signals $\mathbf{y}_{v}$ exhibit a nonlinear relationship with the underlying SCSI, traditional methods for obtaining SCSI  typically require iterative procedures \cite{Citelu20232dSCI}, resulting in high computational complexity. To tackle this challenge and estimate the SCSI efficiently, we  establish a tensor decomposition framework for  location-specific SCSI database construction in the following subsection. 

\subsection{Problem Formulation}
\label{Problem Formulation}
 We aim to extract \( \bar{\tau}_{g, l} \), \( \bar{\theta}_{g, l} \), \( \bar{\varphi}_{g, l} \), and \( \bar{\rho}_{g, l} \) from the data \( \mathbf{H}_{g} \) to construct the SCSI database. Note that the construction of the SCSI database for each grid  is independent and follows the same procedure. Without loss of generality, we present the problem formulation for the \( g \)-th grid below. Using the expression of \( \mathbf{H}_{g} \) in \eqref{HgMultipath}, we first derive a fourth-order tensor \( \boldsymbol{\mathcal{H}}^g \in \mathbb{C}^{N_{\mathrm{d}} \times M_{\mathrm{v}}  \times  M_{\mathrm{h}} \times W }\), with its \( (n, m_{\mathrm{v}}, m_{\mathrm{v}},w) \)-th entry given by \( \left[\mathbf{H}_{g}\right]_{n M+m_{\mathrm{v}}M_{\mathrm{h}}+m_{\mathrm{h}},w } \). 
By comparing \eqref{HgMultipath} to the definition of tensor canonical polyadic decomposition (CPD)\cite{CiteTensorCPD}, it can be readily observed that \( \pmb{\mathcal{H}}_g \) can be represented in a CPD format, i.e.,
\begin{equation}
	\begin{aligned}
	\boldsymbol{\mathcal{H}}_{g}&=\llbracket \mathbf{B}^{(1)}, \mathbf{B}^{(2)},\mathbf{B}^{(3)},\mathbf{B}^{(4)} \rrbracket +\boldsymbol{\mathcal{N}}_{g} \\
	&=\sum_{l=1}^{\bar{L}} \mathbf{b}_{N_{\mathrm{d}}}\left(\bar{\tau}_{g, l}\right) \circ  \mathbf{a}_{\mathrm{v}}(\bar{\theta}_{g, l}) \circ  \mathbf{a}_{\mathrm{h}}( \bar{\varphi}_{g, l}, \bar{\theta}_{g, l})\circ\mathbf{p}_{g,l}+\boldsymbol{\mathcal { N }}_{g},
	\end{aligned}
	\label{CPD format of Hg}
\end{equation}
where $\boldsymbol{\mathcal{N}}_{g}$ is the tensor form of noise and representation error. $\mathbf{B}^{(1)} , \mathbf{B}^{(2)}, \mathbf{B}^{(3)}$ and $\mathbf{B}^{(4)}$ are factor matrices represented by
\begin{equation}
	\begin{aligned}
		& \mathbf{B}^{(1)}=\left[\mathbf{b}_{N_{\mathrm{d}}}\left(\bar{\tau}_{g, 1}\right),\dots,\mathbf{b}_{N_{\mathrm{d}}}\left(\bar{\tau}_{g, \bar{L}}\right) \right]\in \mathbb{C}^{N_{\mathrm{d}} \times \bar{L}},\\
		&\mathbf{B}^{(2)}= \left[ \mathbf{a}_{\mathrm{v}}(\bar{\theta}_{g, 1}),
		\dots,\mathbf{a}_{\mathrm{v}}(\bar{\theta}_{g, \bar{L}}) \right]\in \mathbb{C}^{M_{\mathrm{v}} \times \bar{L}}, \\
		&\mathbf{B}^{(3)}=\left[  \mathbf{a}_{\mathrm{h}}( \bar{\varphi}_{g, 1}, \bar{\theta}_{g, 1}),\dots,\mathbf{a}_{\mathrm{h}}( \bar{\varphi}_{g, \bar{L}}, \bar{\theta}_{g, \bar{L}}) \right]  \in \mathbb{C}^{M_{\mathrm{h}} \times \bar{L}}, \\
		&\mathbf{B}^{(4)}=\left[\mathbf{p}_{g,1},\dots, \mathbf{p}_{g, \bar{L}} \right] \in \mathbb{C}^{W \times \bar{L}}.
	  \label{Factor matries of Hg}
	\end{aligned}
\end{equation}

We aim to estimate the SCSI \( \{ \bar{\tau}_{g, l}, \bar{\theta}_{g, l}, \bar{\varphi}_{g, l}, \bar{\rho}_{g, l} \}_{l=1}^{\bar{L}} \) from the observation tensor \( \boldsymbol{\mathcal{H}}_{g} \), by utilizing the structured CPD format as described in \eqref{CPD format of Hg}-\eqref{Factor matries of Hg}. The SCSI estimation problem can be formulated as the following optimization problem:
\begin{equation}
	\min _{\{ \bar{\tau}_{g, l}, \bar{\theta}_{g, l}, \bar{\varphi}_{g, l}, \bar{\rho}_{g, l} \}_{l=1}^{\bar{L}}}\left\|\boldsymbol{\mathcal{H}}_{g}-\llbracket \mathbf{B}^{(1)}, \mathbf{B}^{(2)}, \mathbf{B}^{(3)}, \mathbf{B}^{(4)} \rrbracket \right\|_F^2.
	\label{CPDecompositionProblem}
\end{equation}
The above problem is essentially a CPD problem, which can be addressed  using  existing tensor decomposition algorithms. 
One of the most widely used algorithms is the alternating least squares (ALS)\cite{CiteTensorALS} method,  which iteratively estimates one factor matrix while keeping the others fixed, updating them alternately. Given the estimated factor matrices \( \{ \mathbf{B}^{(1)}, \mathbf{B}^{(2)}, \mathbf{B}^{(3)}, \mathbf{B}^{(4)} \} \), the SCSI can be derived by leveraging   the underlying manifold structure of  the factor matrices. 

However, the ALS algorithm guarantees the uniqueness of the estimated factor matrices only when the tensor is of low rank. In contrast, the scenario considered in this paper involves a channel with several hundred sub-paths, resulting in a high-rank CPD problem for which the uniqueness condition is no longer satisfied. Since uniqueness is critical for reliable CPD-based decomposition, we first analyze the uniqueness conditions associated with the CPD formulation in equation \eqref{CPDecompositionProblem}. This analysis then serves as the foundation for enhancing the subsequent SCSI estimation algorithm.

\subsection{Extraction of SCSI}
The uniqueness condition of the CPD problem is fundamental to the accurate estimation of SCSI, ensuring that the decomposed factor matrices incorporate the accurate information of channel statistical parameters. A well-known sufficient condition for the uniqueness of CPD  problem \eqref{CPDecompositionProblem}  is as follows\cite{Cite1977kruskalTensor}:

\textit{Lemma 2:} Considering the fourth-order tensor $\boldsymbol{\mathcal{H}}_g$ defined in \eqref{CPD format of Hg}, if the condition $\sum_{i=1}^{4}\mathrm{kr}(\mathbf{B}^{(i)}) \geq 2\bar{L}+3$ is satisfied, then the CPD  of $\boldsymbol{\mathcal{H}}_g$ is guaranteed to be unique. In the general case, the uniqueness condition  simplifies to
\begin{equation}
	\begin{aligned}
	\min \left(N_{\mathrm{d}}, \bar{L}\right)+\min \left(M_{\mathrm{v}}, \bar{L}\right)+\min \left(M_{\mathrm{h}}, \bar{L}\right) +&\min \left(W, \bar{L}\right)\\
	&  \geq 2\bar{L}+3.
	\end{aligned}
		\label{kruscal Condition}
\end{equation}

In our system,  since \( M_v \), \( M_h \), and \( W \) are much smaller than \( \bar{L} \),  the condition $		\mathrm{kr}\left(\mathbf{B}^{(2)}\right)+		\mathrm{kr}\left( \mathbf{B}^{(3)} \right)+		\mathrm{kr}\left(\mathbf{B}^{(4)} \right)<\bar{L}$ holds. Therefore, the Kruskal condition  cannot be satisfied.

To address this issue, the structural properties within the tensor must be exploited to relax the uniqueness condition. Note that the factor matrices \( \mathbf{B}^{(1)} \), \( \mathbf{B}^{(2)} \), and \(\mathbf{B}^{(3)} \) are all Vandermonde matrices where the generators are  $\{z_{1,l} = e^{-\bar{\jmath} 2 \pi  \Delta f \bar{\tau}_{g, l}}\}_{l=1}^{\bar{L}},\{z_{2,l} =  e^{-\bar{\jmath} \pi \cos  \bar{\theta}_{g, l}}\}_{l=1}^{\bar{L}}$ and $\{z_{3,l} =  e^{-\bar{\jmath} \pi \sin  \bar{\theta}_{g, l} \cos \bar{\varphi}_{g, l} }\}_{l=1}^{\bar{L}}$ respectively. Utilizing this structural information, the following relaxed uniqueness condition can be derived\cite{CiteVSDTensorDecompositionMethod,CiteTensorCPDBlindSignalSeparation}.

\textit{Lemma 3:} Considering the fourth-order tensor $\boldsymbol{\mathcal{H}}_g$ defined in \eqref{CPD format of Hg}, where $\mathbf{B}^{(1)}, \mathbf{B}^{(2)}, \mathbf{B}^{(3)}$ are Vandermonde matrices with generators $\{z_{1, l}\}_{l=1}^{\bar{L}},\{z_{2, l}\}_{l=1}^{\bar{L}},\{z_{3, l}\}_{l=1}^{\bar{L}}$.   Select the smoothing parameters \( (K_s, L_s), s=1,2,3 \) subject to \( K_1 + L_1 = N_{\mathrm{d}} + 1 \), \( K_2 + L_2 = M_{\mathrm{v}} + 1 \), and \( K_3 + L_3 = M_{\mathrm{h}} + 1 \) for spatial smoothing. If
\begin{gather}
\begin{aligned}
	& z_{1, i} \neq z_{1, j}, \forall i \neq j,\\
	& \mathrm{r}\left(\mathbf{B}^{\left(K_1-1,1\right)} \odot \mathbf{B}^{\left(K_2, 2\right)} \odot \mathbf{B}^{\left(K_3, 3\right)}\right)=\bar{L}, \\
	& \mathrm{r} \left(\mathbf{B}^{\left(L_1, 1\right)} \odot \mathbf{B}^{\left(L_2, 2\right)} \odot \mathbf{B}^{\left(L_3, 3\right)} \odot \mathbf{B}^{(4)}\right)=\bar{L},
\end{aligned}
\label{VandermondeUniquenessCondition}
\end{gather}
then the CPD of $\boldsymbol{\mathcal{H}}_g$ is unique. Specifically, $\mathbf{B}^{\left(K_1-1,1\right)} \triangleq [\mathbf{B}^{(1)}]_{1: K_1-1,:}$ denotes the first $K_1-1$ rows of $\mathbf{B}^{(1)}$. In general, condition \eqref{VandermondeUniquenessCondition} simplifies to
\begin{gather}
\min  \left(\left(K_1-1\right) K_2 K_3, L_1 L_2 L_3 I_4\right) \geq \bar{L}.
\label{VandermondeUniquenessCondition1}
\end{gather}
The relaxed uniqueness condition \eqref{VandermondeUniquenessCondition1} can be guaranteed by appropriately choosing parameters \( (K_s, L_s), s=1,2,3 \).

Based on Lemma 3, the high-rank CPD problem in our work can be solved by exploiting the Vandermonde structure of the factor matrices.
Specifically, we leverage the method in \cite{CiteVSDTensorDecompositionMethod}  and \cite{CiteTensorCPDBlindSignalSeparation} to estimate the SCSI of grid \( g \) using \( \boldsymbol{\mathcal{H}}_{g} \). We define the smoothing parameters \( (K_s,L_s), s = 1, 2, 3 \), which satisfy the conditions outlined in Lemma 3.  Consider the matricization $\mathbf{X}^{[3]}$ of the tensor $\boldsymbol{\mathcal{H}}_{g}$\cite{CiteTensorCPDBlindSignalSeparation} as follows:
\begin{equation}
	\begin{aligned}
		&\mathbf{X}^{[3]}   \\
		&\triangleq \small \left[\begin{array}{ccc@{}c}
			\left[\boldsymbol{\mathcal{H}}_{g}\right]_{1, 1, 1, 1} & 	\left[\boldsymbol{\mathcal{H}}_{g}\right]_{1, 1, 1, 1} & \cdots & 	\left[\boldsymbol{\mathcal{H}}_{g}\right]_{1, 1, 1, W} \\
			\left[\boldsymbol{\mathcal{H}}_{g}\right]_{1, 1, 2, 1} & 	\left[\boldsymbol{\mathcal{H}}_{g}\right]_{1, 1, 2, 1} & \cdots & 	\left[\boldsymbol{\mathcal{H}}_{g}\right]_{1, 1, 2, W}  \\
			\vdots & \vdots & \ddots & \vdots \\
			\left[\boldsymbol{\mathcal{H}}_{g}\right]_{N_{\mathrm{d}}, M_{\mathrm{v}}, M_{\mathrm{h}}, 1} & 	\left[\boldsymbol{\mathcal{H}}_{g}\right]_{N_{\mathrm{d}}, M_{\mathrm{v}}, M_{\mathrm{h}}, 2} & \cdots & 	\left[\boldsymbol{\mathcal{H}}_{g}\right]_{N_{\mathrm{d}}, M_{\mathrm{v}}, M_{\mathrm{h}}, W}
		\end{array}\right] \\
		& =\left(\mathbf{B}^{(1)} \odot \mathbf{B}^{(2)} \odot \mathbf{B}^{(3)} \right) \mathbf{P}^{T}_{g}+\mathbf{N}^{[3]},
		\label{MatricizationOfH_g}
	\end{aligned}
\end{equation}
where $\mathbf{N}^{[3]} \in \mathbb{C}^{N_{\mathrm{d}}M_{\mathrm{v}} M_{\mathrm{h}} \times W}$ is the corresponding noise matrix. Since \( \mathbf{B}^{(1)}, \mathbf{B}^{(2)} \), and \( \mathbf{B}^{(3)} \) exhibit a Vandermonde structure, the dimension of \( \mathbf{X}^{[3]} \) can be  expanded   by exploiting the spatial smoothing technique. We perform spatial smoothing on $\mathbf{X}^{[3]}$, yielding
\begin{equation}
	\begin{aligned}
		\mathbf{X}_{\mathcal{S}} & \triangleq {\left[\begin{array}{lllll}
				\mathbf{J}_{1,1,1}\mathbf{X}^{[3]} &\cdots& \mathbf{J}_{1,1,L_3} \mathbf{X}^{[3]}  \quad \cdots  & \mathbf{J}_{1,2,1} \mathbf{X}^{[3]}
			\end{array}\right.}\\
		& \left.\begin{array}{lllll}
				 & \qquad \qquad  \cdots &\mathbf{J}_{1,2,L_3} \mathbf{X}^{[3]}     & \quad \cdots\quad  \mathbf{J}_{1,L_2,L_3} \mathbf{X}^{[3]} 
			\end{array}\right.\\
		& \left.\begin{array}{l@{}l@{}l@{}l@{}l}
				& \quad \qquad  \qquad \qquad \cdots \quad &\mathbf{J}_{L_1,1,1} \mathbf{X}^{[3]}     & \quad \cdots\quad  \mathbf{J}_{L_1,L_2,L_3} \mathbf{X}^{[3]} 
			\end{array}\right]\\
		&=  \left(\mathbf{B}^{\left(K_1, 1\right)} \odot  \mathbf{B}^{\left(K_2, 2\right)} \odot \mathbf{B}^{\left(K_3, 3\right)}\right) \\
		& \cdot \left(\mathbf{B}^{\left(L_1, 1\right)} \odot  \mathbf{B}^{\left(L_2, 2\right)} \odot \mathbf{B}^{\left(L_3, 3\right)} \odot \mathbf{B}^{(4)}\right)^T+\mathbf{N}_{\mathcal{S}},
		\label{eq:SpatialSmoothing}
	\end{aligned}
\end{equation}
where $\mathbf{N}_{\mathcal{S}}  \in \mathbb{C}^{K_1 K_2 K_3 \times L_1 L_2 L_3 W}$ is the corresponding noise matrix;  $\mathbf{J}_{l_1, l_2, l_3}$ is the selection matrix \cite{CiteTensorSpatialSmoothingSelectionMatrix}, given by
\begin{equation}
	\begin{aligned}
		&\mathbf{J}_{l_1, l_2, l_3} \triangleq {\left[\begin{array}{lll}
				\mathbf{0}_{K_1 \times\left(l_1-1\right)} & \mathbf{I}_{K_1} & \mathbf{0}_{K_1 \times\left(L_1-l_1\right)}
			\end{array}\right]}\otimes \\
		& \qquad {\left[\begin{array}{lll}
				\mathbf{0}_{K_2 \times\left(l_2-1\right)}& \mathbf{I}_{K_2} & \mathbf{0}_{K_2 \times\left(L_2-l_2\right)}
			\end{array}\right]} \otimes \\
		&\qquad \left[\begin{array}{lll}
			\mathbf{0}_{K_3 \times\left(l_3-1\right)} & \mathbf{I}_{K_3} & \mathbf{0}_{K_3 \times\left(L_3-l_3\right)}
		\end{array}\right].
	\end{aligned}
\end{equation}

Subsequently, the  truncated singular value decomposition is performed on \( \mathbf{X}_{\mathcal{S}} \), i.e.,
\begin{equation}
	\operatorname{SVD}\left(\mathbf{X}_{\mathcal{S}} \right)=\mathbf{U} \boldsymbol{\Sigma} \mathbf{V}^{\mathrm{H}},
	\label{Eq: InitialSVD}
\end{equation}
where $\mathbf{U} \in \mathbb{C}^{K_1 K_2 K_3 \times \bar{L}},  \boldsymbol{\Sigma} \in \mathbb{C}^{\bar{L}\times \bar{L}} $ and $ \mathbf{V} \in \mathbb{C}^{ L_1 L_2 L_3 W \times \bar{L}} $. Using the minimum description length (MDL) criterion \cite{CiteTensorMDL}, \( \bar{L} \) can be estimated.

Omitting the noise and based on \eqref{VandermondeUniquenessCondition}, there exists a non-singular matrix \( \mathbf{M} \in \mathbb{C}^{\bar{L} \times \bar{L}} \) such that
\begin{equation}
	\begin{aligned}
			\mathbf{U M} &=\mathbf{B}^{\left(K_{1}, 1\right)} \odot \mathbf{B}^{\left(K_{2}, 2\right)} \odot \mathbf{B}^{\left(K_{3}, 3\right)}, \\
		\mathbf{V}^* \Sigma \mathbf{N}  &=\mathbf{B}^{\left(L_1, 1\right)} \odot  \mathbf{B}^{\left(L_2, 2\right)} \odot \mathbf{B}^{\left(L_3, 3\right)} \odot \mathbf{B}^{(4)}, \mathbf{N}=\mathbf{M}^{-T}.
	\end{aligned}
	 	\label{Eq:SVDUSigmaV}
\end{equation}
The above equation implies that
 \begin{equation}
 	\begin{aligned}
 		& \mathbf{U}_1 \mathbf{M}=\underline{\mathbf{B}}^{\left(K_{1}, 1\right)} \odot \mathbf{B}^{\left(K_{2}, 2\right)} \odot \mathbf{B}^{\left(K_{3}, 3\right)}, \\
 		& \mathbf{U}_2 \mathbf{M}=\overline{\mathbf{B}}^{\left(K_{1}, 1\right)} \odot \mathbf{B}^{\left(K_{2}, 2\right)} \odot \mathbf{B}^{\left(K_{3}, 3\right)},
 	\end{aligned}
 	\label{Eq:SVDFactorU1U2}
 \end{equation}
where $\underline{\mathbf{B}}^{\left(K_{1}, 1\right)}$ and $\overline{\mathbf{B}}^{\left(K_{1}, 1\right)}$ are obtained by deleting the bottom and top
 row of $\mathbf{B}^{\left(K_{1}, 1\right)}$ respectively, i.e., $\underline{\mathbf{B}}^{\left(K_{1}, 1\right)}=[\mathbf{B}^{\left(K_{1}, 1\right)}]_{1:K_1-1,:}$ ,  $\overline{\mathbf{B}}^{\left(K_{1}, 1\right)}=[\mathbf{B}^{\left(K_{1}, 1\right)}]_{2:K_1,:}$.  The expressions for \( \mathbf{U}_1 \) and \( \mathbf{U}_2 \) are
\begin{equation}
	\begin{aligned}
		& \mathbf{U}_1=[\mathbf{U}]_{1:\left(K_1-1\right) K_2 K_3, :} , \\
		& \mathbf{U}_2=[\mathbf{U}]_{1+K_2 K_3: K_1 K_2 K_3, :}.
	\end{aligned}
	\label{Eq:SVDU1U2}
\end{equation}

Due to the Vandermonde structure of \( \mathbf{B}^{(1)} \), we have
\begin{equation}
	\begin{aligned}
	&\left(\underline{\mathbf{B}}^{\left(K_{1}, 1\right)} \odot \mathbf{B}^{\left(K_{2}, 2\right)} \odot \mathbf{B}^{\left(K_{3}, 3\right)}\right)\mathbf{Z}_{1}\\
	&\qquad \qquad  =\overline{\mathbf{B}}^{\left(K_{1}, 1\right)} \odot \mathbf{B}^{\left(K_{2}, 2\right)} \odot \mathbf{B}^{\left(K_{3}, 3\right)},
		\end{aligned}
	\label{Eq:FactorB1B2B3}
\end{equation}
where $\mathbf{Z}_{1}=\operatorname{diag}\left(\left[z_{1,1}, \ldots, z_{1,\bar{L}}\right]\right)$.

By merging \eqref{Eq:SVDFactorU1U2}-\eqref{Eq:FactorB1B2B3}, the following equation is obtained:
\begin{equation}
	\mathbf{U}_1 \mathbf{M} \mathbf{Z}_{1}= \mathbf{U}_2 \mathbf{M}.
	\label{Eq:U1U2Z1}
\end{equation}

From \eqref{Eq:U1U2Z1}, we have \( \mathbf{U}_2 = \mathbf{U}_1 \widehat{\mathbf{Z}}_1 \), where \( \widehat{\mathbf{Z}}_1 = \mathbf{M} \mathbf{Z}_1 \mathbf{M}^{-1} \). Since \( \underline{\mathbf{B}}^{(K_1, 1)} \odot \mathbf{B}^{(K_2, 2)} \odot \mathbf{B}^{(K_3, 3)} \) has full column rank, it follows that \( \mathbf{U}_1 \) and \( \mathbf{U}_2 \) also have full column rank. Therefore, we obtain \( \widehat{\mathbf{Z}}_1 = \mathbf{U}_1^{\dagger} \mathbf{U}_2 \). From the eigenvalue decomposition (EVD), \( \mathbf{U}_1^{\dagger} \mathbf{U}_2  = \mathbf{M} \mathbf{Z}_1 \mathbf{M}^{-1} \), the Vandermonde generator  set \( \{ z_{1,l} \}_{l=1}^{\bar{L}} \)  of $\mathbf{B}^{(1)}$ is derived, i.e.,
\begin{equation}
	\left\{z_{1, l}\right\}_{l=1}^{\bar{L}}=\operatorname{diag}(\mathbf{Z}_{1}), z_{1, l}=\frac{z_{1, l}}{\left|z_{1, l}\right|}, l=1, \ldots, \bar{L}.
	\label{Cal:NormalizedZ_1}
\end{equation}

Then,  we can reconstruct $\mathbf{B}^{(1)}$ with $\left\{z_{1, l}\right\}_{l=1}^{\bar{L}}$ based on  \eqref{Factor matries of Hg}. The next step is to find $\mathbf{B}^{(2)}$. Note that
\begin{equation}
	\begin{aligned}
	\left(\frac{(\mathbf{b}_l^{(K_1, 1) })^{H}}{\mathbf{b}_l^{(K_1, 1) H} \mathbf{b}_l^{(K_1,1)}} \otimes \mathbf{I}_{K_2 K_3}\right)&\left(\mathbf{b}_l^{(K_1,1)} \otimes \mathbf{b}_l^{(K_2,2)} \otimes \mathbf{b}_l^{(K_3,3)}\right)\\
	& =\mathbf{b}_l^{(K_2,2)} \otimes \mathbf{b}_l^{(K_3,3)},
	\end{aligned}
\end{equation}
where $\mathbf{b}_{l}^{\left(K_s, s\right)}$ is the $l$-th column of $\mathbf{B}^{\left(K_{s}, s\right)}$($s=1,2,3$). Thus, by leveraging \eqref{Eq:SVDUSigmaV}, $\mathbf{b}_l^{(K_2,2)} \otimes \mathbf{b}_l^{(K_3,3)}$ is obtained as
\begin{equation}
\mathbf{b}_l^{(K_2,2)} \otimes \mathbf{b}_l^{(K_3,3)}=\left(\frac{\left(\mathbf{b}_l^{\left(K_1, 1\right)} \right)^{\mathrm{H}}} {\left(\mathbf{b}_l^{\left(K_1, 1\right)} \right)^{\mathrm{H}} \mathbf{b}_l^{\left(K_1, 1\right)} } \otimes \mathbf{I}_{K_2 K_3}\right) \mathbf{U} \mathbf{m}_l.
\end{equation}

 Then,  the second Vandermonde generator set $\left\{z_{2, l}\right\}_{l=1}^{\bar{L}}$ of $\mathbf{B}^{(2)}$ is determined. Since $\left( \underline{\mathbf{B}}^{\left(K_{2}, 2\right)} \odot \mathbf{B}^{\left(K_{3}, 3\right)}\right)\mathbf{Z}_{2}= \overline{\mathbf{B}}^{\left(K_{2}, 2\right)} \odot \mathbf{B}^{\left(K_{3}, 3\right)}$, where $\mathbf{Z}_{2}=\operatorname{diag}\left(\left[z_{2,1}, \ldots, z_{2,\bar{L}}\right]\right)$, we have
\begin{equation}
	\begin{aligned}
	 z_{2, l}&=\left(\mathbf{b}_l^{(K_2,2)} \otimes \mathbf{b}_l^{(K_3,3)} \right)_{\left(1:\left(K_2-1\right) K_3, 1\right)}^{\dagger}\\
	 & \cdot \left(\mathbf{b}_l^{(K_2,2)} \otimes \mathbf{b}_l^{(K_3,3)} \right)_{\left(K_3+1: K_2 K_3, 1\right)}.
	 \end{aligned}
	 \label{cal:NormailizedZ2}
\end{equation}

Similarly, we reconstruct $\mathbf{B}^{(2)}$ with $\left\{z_{2, l}\right\}_{l=1}^{\bar{L}}$. So far, we have obtained $\left\{\mathbf{B}^{(1)}, \mathbf{B}^{(2)}\right\}$. $l$-th vector of $\mathbf{B}^{(3)}$ can be derived in the similar way, i.e.,
\begin{equation}
	\begin{aligned}
\mathbf{b}_{l}^{\left(K_3, 3\right)}&=\left(\frac{\left(\mathbf{b}_l^{\left(K_2, 2\right)} \right)^{\mathrm{H}}} {\left(\mathbf{b}_l^{\left(K_2, 2\right)} \right)^{\mathrm{H}} \mathbf{b}_l^{\left(K_2, 2\right)} } \otimes \mathbf{I}_{K_3}\right) \\
&\quad \cdot \left(\frac{\left(\mathbf{b}_l^{\left(K_1, 1\right)} \right)^{\mathrm{H}}} {\left(\mathbf{b}_l^{\left(K_1, 1\right)} \right)^{\mathrm{H}} \mathbf{b}_l^{\left(K_1, 1\right)} }\otimes \mathbf{I}_{K_2 K_3}\right) \mathbf{U m}_l.
\end{aligned}
	 \label{cal:NormailizedZ3}
\end{equation}

Considering the Vandermonde structure of \( \mathbf{B}^{(K_3, 3)} \),  the generators of \( \mathbf{B}^{(3)} \) can be expressed as \( z_{3,l} = \left( \underline{\mathbf{b}_l}^{(K_3, 3)} \right)^{\dagger} \overline{\mathbf{b}}_l^{(K_3, 3)} \). According to \eqref{Eq:SVDUSigmaV}, $l$-th vector of $\mathbf{B}^{(4)}$ can be derived as
\begin{equation}
	\begin{aligned}
	\mathbf{b}_l^{(4)}&=\frac{\left(\mathbf{b}_l^{\left(L_1, 1\right)} \right)^{\mathrm{H}}} {\left(\mathbf{b}_l^{\left(L_1, 1\right)} \right)^{\mathrm{H}} \mathbf{b}_l^{\left(L_1, 1\right)} } \otimes\frac{\left(\mathbf{b}_l^{\left(L_2, 2\right)} \right)^{\mathrm{H}}} {\left(\mathbf{b}_l^{\left(L_2, 2\right)} \right)^{\mathrm{H}} \mathbf{b}_l^{\left(L_2, 2\right)} } \otimes \\
	&\frac{\left(\mathbf{b}_l^{\left(L_3, 3\right)} \right)^{\mathrm{H}}} {\left(\mathbf{b}_l^{\left(L_3, 3\right)} \right)^{\mathrm{H}} \mathbf{b}_l^{\left(L_3, 3\right)} }\otimes \mathbf{I}_{W} \cdot \mathbf{V}^* \Sigma \mathbf{n}_l.
	\end{aligned}
	\label{Eq: Compute bl4}
\end{equation}

Upon completing the tensor decomposition, we proceed to estimate the SCSI. Specifically, the SCSI \( \{\bar{\tau}_{g,l}, \bar{\theta}_{g,l}, \bar{\varphi}_{g,l}, \bar{\rho}_{g,l}\}_{l=1}^{\bar{L}} \) can be estimated using the generators \( \{ z_{1,l}, z_{2,l}, z_{3,l} \}_{l=1}^{\bar{L}} \) and the factor matrix \( \mathbf{B}^{(4)} \) as follows:
\begin{subequations}
\begin{align}
\bar{\tau}_{g,l}&\triangleq -\frac{1}{2 \pi \Delta f} \angle z_{1, l}, l=1, \ldots, \bar{L}, \label{Eq: CalTauViaZ1}\\
\bar{\theta}_{g,l} &\triangleq \arccos \left(-\frac{1}{ \pi } \angle z_{2, l}\right),l=1, \ldots, \bar{L}, \label{Eq: CalThetaViaZ2} \\
\bar{\varphi}_{g, l} &\triangleq \arccos \left(- \frac{1}{ \pi \sin  \bar{\theta}_{g, l}} \angle z_{3, l} \right),  l=1, \ldots, \bar{L}, \label{Eq: CalvarphiViaZ3} \\
\bar{\rho}_{l,g}&\triangleq\frac{1}{W}\| 	\mathbf{b}_l^{(4)}\|^2_2,  l=1, \ldots, \bar{L}.
\label{Eq: CalRhoViaB4}
\end{align}
\end{subequations}
The VSTD-based SCSI database construction algorithm is summarized in Algorithm \ref{TensorBasedSCSIConstruction}.

\begin{algorithm}[!t]
	\caption{VSTD-Based SCSI Database Construction Algorithm for Grid $g$} 
	\label{TensorBasedSCSIConstruction}
	\hspace*{0in} {\bf {Input:}} $\boldsymbol{\mathcal{H}}_{g} \in \mathbb{C}^{N_{\mathrm{d}} \times M_{\mathrm{v}}  \times  M_{\mathrm{h}} \times W }$
	\begin{algorithmic}[1]
         \State
         Compute  $\mathbf{X}_{\mathcal{S}}$  using \eqref{MatricizationOfH_g} and \eqref{eq:SpatialSmoothing}.
		 \State
		 Compute  the SVD of $\mathbf{X}_{\mathcal{S}}$  using \eqref{Eq: InitialSVD}.
		 \State
		 Compute EVD as 	$\mathbf{U}_1^{\dagger}\mathbf{U}_2 =  \mathbf{M} \mathbf{Z}_{1} \mathbf{M}^{-1}$.		
		 \State
		Estimate the normalized generators 	\( \{ z_{1,l}, z_{2,l}, z_{3,l} \}_{l=1}^{\bar{L}} \)		 using 	\eqref{Eq:U1U2Z1}-\eqref{cal:NormailizedZ3}.
		\State
		Reconstruct $ \left\{\mathbf{B}^{(1)}, \mathbf{B}^{(2)},  \mathbf{B}^{(3)} \right\}$ using 	\( \{ z_{1,l}, z_{2,l}, z_{3,l} \}_{l=1}^{\bar{L}} \).
		\State
		Compute the factor matrices $ \mathbf{B}^{(4)}$ using \eqref{Eq: Compute bl4}.
		\State
        Compute $\left\{\bar{\tau}_{g,l}, \bar{\theta}_{g,l}, \bar{\varphi}_{g, l}, \bar{\rho}_{l,g}\right\}_{l=1}^{\bar{L}}$ via \eqref{Eq: CalTauViaZ1} -\eqref{Eq: CalRhoViaB4}.
	\end{algorithmic}
	\hspace*{0in} {\bf {Output:}}$\{\bar{\tau}_{g, l}, \bar{\theta}_{g, l}, \bar{\varphi}_{g, l}, \bar{\rho}_{g, l}\}_{l=1}^{\bar{L}}$.
\end{algorithm}

\label{SCSI-Assited Multi-user channel estimation}
\section{Simulation Results}
\label{sec_Simulation}
\subsection{Simulation Configuration}
In this section, we present simulation results to evaluate the performance of the proposed algorithms. To generate the channels for simulations in the 3GPP 38.901 urban macro (UMa) line-of-sight (LOS) scenario, we utilize QuaDRiGa\cite{Cite2014Quadriga}, which is capable of generating massive MIMO-OFDM channels that consistent with the 3GPP 38.901 specifications\cite{Cite3GPP38.901Protocol}. Each channel consists of $34$ clusters, comprising  $221$ subpaths. Furthermore, the channel parameters vary at different positions in accordance with the spatial consistency procedure \cite{Cite3GPP38.901Protocol}. The basic simulation parameters are presented in Table \ref{tab:system parameters}. In addition,  the SNR during channel estimation and the location-specific SCSI database construction are denoted as $\text{SNR}_\mathrm{CE}$ and $\text{SNR}_\mathrm{SC}$, respectively.

\begin{table}[!t]
	\small
	\centering
	\caption{Basic System Parameters}
	\label{tab:system parameters}
	\begin{tabular}{c|c}
		\hline
		\textbf{System Parameters} & \textbf{Value}  \\ \hline 
		Centering frequency $ f_{\text{c}} $ & 6.7 GHz \\ 
		Bandwidth $ B $  & 100 MHz  \\ 
		FFT size $ N_{\text{FFT}} $ & 4096 \\ 
		Number of subcarriers for transmission  $ N_{\mathrm{c}} $ & 816 \\ 
		Subcarrier spacing $ \Delta f $ & 30 kHz \\ 
		Number of  CDM group $G$ & 3 \\ 
		Number of  OFDM symbols for pilots $ T_{\mathrm{p}} $ & 2 \\ 
		Number of subcarriers each CDM  group $ N $ & 272 \\ 
		Number of UE $K $ & 24 \\ 
		Number of spatial sampling points for each grid $W $ & 10 \\ 
		Number of BS antennas $ [M_{\text{v}},M_{\text{h}}] $ & [4,16]  \\ 
		Height of BS $ h_\text{BS} $ & 25 m \\ 
		Height of UE $ h_\text{UE}  $ & 1.5m  \\ 
		Velocity of UE $v$& 0.1 km/h  \\ 
		Delay spread & 300ns  \\ 
		 Shape parameter of the Kaiser window & 3.95\\
		 \hline 
	\end{tabular}
\end{table}

\subsection{ Benchmarks and Performance Metrics}
To demonstrate the superiority of the proposed scheme, we compare our schemes with the following state-of-the-art algorithms:
\begin{itemize}
\item OMP without SCSI Database\cite{CiteOMPBaseLine}: The orthogonal matching pursuit  (OMP) algorithm  iteratively identifies the first \(L_{t,k}\) optimal matches for \(\tau_{l,t,k}\), \(\theta_{l,t,k}\), and \(\varphi_{l,t,k}\) based on a delay-angular domain dictionary  from the   trivial approach results.  The channel is subsequently reconstructed using the estimated parameters.
		
	\item VSD without SCSI Database\cite{CiteVSDTensorDecompositionMethod}: The Vandermonde structured decomposition (VSD) algorithm   employs the tensor decomposition to estimate the factor matrices  by leveraging the trivial approach results and the inherent Vandermonde structure of the channel.  Subsequently, the channel is reconstructed using the obtained factor matrices.

	\item EM-AMP without SCSI Database\cite{Citevila2013expectationSCSI}: This algorithm assumes an unknown Bernoulli-Gaussian  prior, learns the hyper-parameters through EM algorithm, and estimates the frequency-space domain channels of all users via approximate message passing (AMP) based on the trivial approach results.
	
	\item SA-BCE with SOMP-based SCSl Database: This algorithm leverages the SCSI database constructed via the  simultaneous orthogonal matching pursuit (SOMP) algorithm for SA-BCE. The SOMP algorithm\cite{CiteSOMPBaseLine} extracts the SCSI under signal model \eqref{HgMultipath}, taking into account the common statistical properties of the channel across different spatial sampling points.

\end{itemize}

To evaluate the performance of  SCSI database construction, we use the  mean square error (MSE)  of location-specific SCSI-assisted MMSE estimator  as the performance metric, i.e., 
\begin{equation}
	\mathcal{L}_{\mathrm{SCSI}}=\frac{\mathcal{E}_{\mathrm{f}}+\mathcal{E}_{\mathrm{s}}}{2},
	\label{Accuracy of SCSI}
\end{equation}
where \( \mathcal{E}_{\mathrm{f}} \) and \( \mathcal{E}_{\mathrm{s}} \) denote the MSE of the frequency-domain MMSE channel estimator and antenna-domain MMSE channel estimator, respectively, which can be expressed as
\begin{equation}
	\begin{aligned}
		&\mathcal{E}_{\mathrm{f}}=\frac{1}{N_d}\operatorname{tr}\left(\mathbf{R}_{\mathrm{f}}-\mathbf{R}_{\mathrm{f}} \tilde{\mathbf{R}}_{\mathrm{f}}\left(\tilde{\mathbf{R}}_{\mathrm{f}}+\sigma^2 \mathbf{I}_{N}\right)^{-1}\right),\\
		&\mathcal{E}_{\mathrm{s}}=\frac{1}{M}\operatorname{tr}\left(\mathbf{R}_{\mathrm{s}}-\mathbf{R}_{\mathrm{s}} \tilde{\mathbf{R}}_{\mathrm{s}}\left(\tilde{\mathbf{R}}_{\mathrm{s}}+\sigma^2 \mathbf{I}_{M}\right)^{-1}\right),
	\end{aligned}
\end{equation}
where \( \mathbf{R}_{\mathrm{f}} \in \mathbb{C}^{N_\mathrm{c}\times N_\mathrm{c}} \) and \( \mathbf{R}_{\mathrm{s}}  \in \mathbb{C}^{M\times M}\) denote the ideal frequency-domain and antenna-domain channel correlation matrices, respectively,  \( \tilde{\mathbf{R}}_{\mathrm{f}} \in \mathbb{C}^{N_\mathrm{c}\times N_\mathrm{c}}\) and \( \tilde{\mathbf{R}}_{\mathrm{s}}   \in \mathbb{C}^{M\times M}\) represent the frequency-domain and antenna-domain channel covariance matrices obtained using the SCSI from location-specific SCSI database.

To assess the performance of channel estimation, we introduce the  normalized mean squared error (NMSE) as the evaluation metric, i.e.,
\begin{equation}
	\mathrm{NMSE}=\frac{1}{T_pK} \sum_{k=1}^{K}\sum_{t=1}^{T_p} 10 \log _{10}\left(\frac{\|\hat{\mathbf{H}}_{k}(t)-\mathbf{H}_{k}(t)\|_{\mathrm{F}}^2}{\|\mathbf{H}_{k}(t)\|_{\mathrm{F}}^2}\right),
\end{equation}
where $\mathbf{H}_{k}(t) \in \mathbb{C}^{N_\mathrm{c}\times M}$ and $\hat{\mathbf{H}}_{k}(t)\in \mathbb{C}^{N_\mathrm{c}\times M}$ represent the actual and estimated  frequency-space domain channel for the $k$-th user at the $t$-th symbol duration.

\subsection{Simulation Results}
\begin{figure}[!t]
	\centering
	\includegraphics[width=0.485\textwidth]{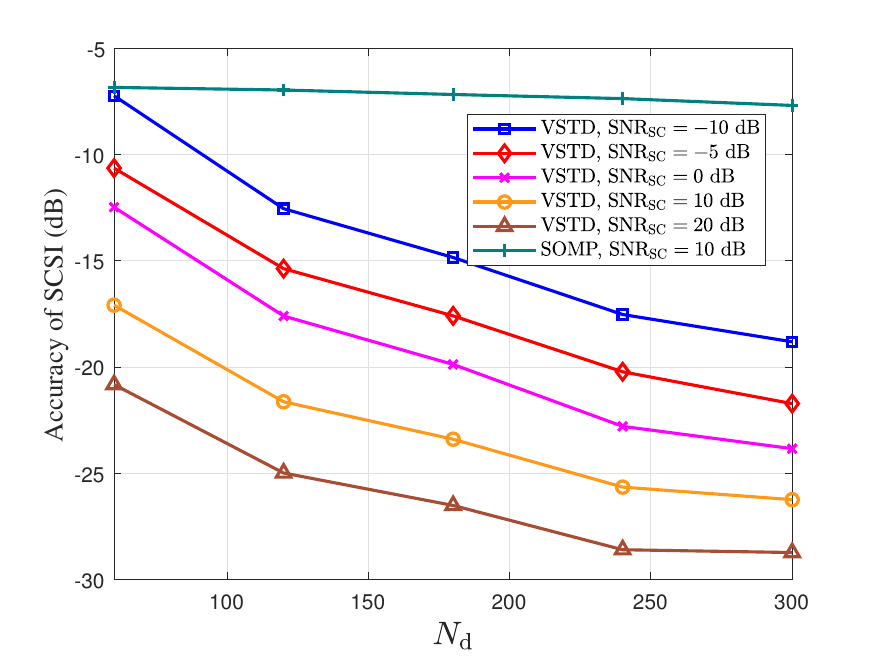}
	\centering
	\caption{SCSI accuracy versus $N_{\mathrm{d}}$ ($d =2$m, $\sigma^2 = 10^{-3}$ ).}
	\label{fig:fig_CKMAccuracyd2mVerusNdDifferentSNR}
\end{figure}

The SCSI accuracy of the proposed VSTD-based algorithm compared to the SOMP-based benchmark under varying SNR and \( N_{\mathrm{d}} \) is depicted in Fig. \ref{fig:fig_CKMAccuracyd2mVerusNdDifferentSNR}. It is evident that the proposed algorithm consistently enhances SCSI accuracy with increases in both \( N_{\mathrm{d}} \) and $\text{SNR}_\mathrm{SC}$ . Across the entire evaluated range of \( N_{\mathrm{d}} \), the VSTD-based SCSI database construction algorithm demonstrates substantial superiority over the SOMP-based baseline.  To quantify, at $ N_{\mathrm{d}} = 180 $ and an $\text{SNR}_\mathrm{SC}$ of $10$ dB, the proposed algorithm realizes a SCSI accuracy of $-23.5$ dB, which is $16.3$ dB superior to the $-7.2$ dB achieved by the SOMP-based approach.

\begin{figure}[t]
	\centering
	\includegraphics[width=0.485\textwidth]{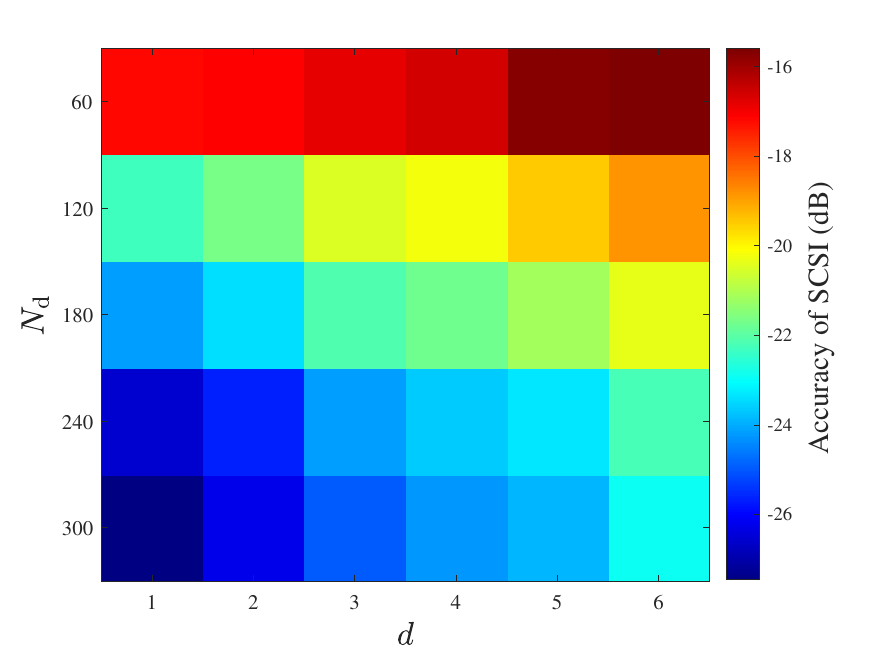}
	\caption{The SCSI accuracy of the VSTD-based  SCSI database at different $N_{\mathrm{d}}$ and $d$ ($\text{SNR}_\mathrm{SC}$ $=10$ dB, $\sigma^2 = 10^{-3}$).}
	\label{fig:CKMAccuracyVerusNdDifferentd}
\end{figure}

 Fig. \ref{fig:CKMAccuracyVerusNdDifferentd} presents the SCSI accuracy of the proposed algorithm as a function of the grid size \( d \) and the number of subcarriers \( N_{\mathrm{d}} \). The proposed algorithm exhibits an improvement in SCSI accuracy with an increase in \( N_{\mathrm{d}} \) or a decrease in \( d \). Notably, to maintain a consistent SCSI accuracy of $-23.4$ dB, an increase in the grid size \( d \) from $2 \, \text{m}$ to $5 \, \text{m}$ necessitates a substantial increase in \( N_{\mathrm{d}} \), specifically from $120$ to $240$. This observation highlights a critical trade-off: while smaller grid sizes enhance SCSI accuracy, they lead to finer grid partitioning and consequently increase the storage overhead for the SCSI database. Thus, a judicious selection of \( d \) is imperative to achieve an optimal balance between accuracy and storage efficiency.

\begin{figure}[!t]
	\centering
	\includegraphics[width=0.485\textwidth]{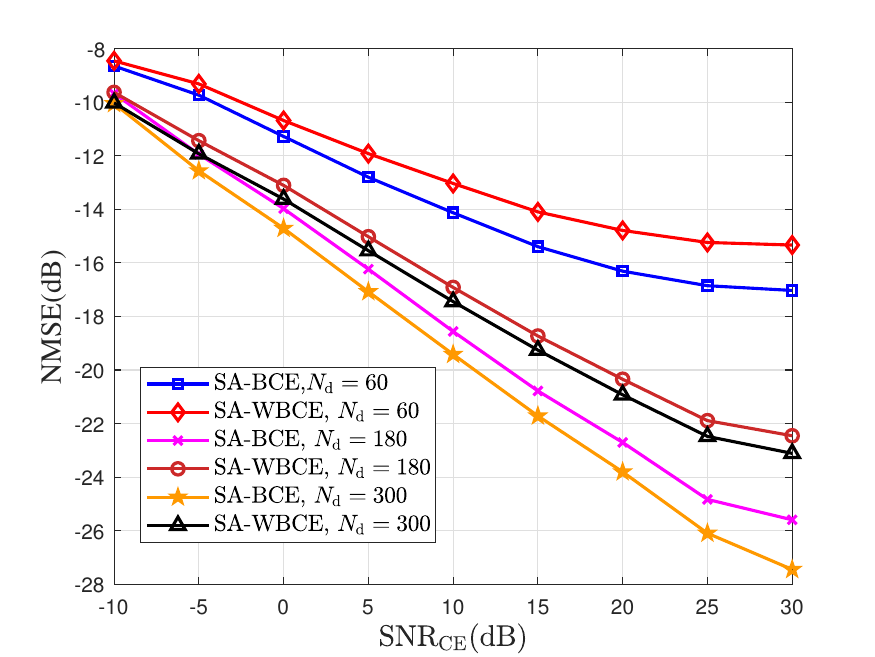}
	\caption{The NMSE of the proposed method with different $N_{\mathrm{d}}$ in SCSI database construction ($\text{SNR}_\mathrm{SC}$ $ = 10$ dB, $d=2$m). }
	\label{fig:CENMSEVerusSNRDifferentNd}
\end{figure}

To examine how the channel estimation performance of the proposed method varies with the number of subcarriers used for location-specific SCSI acquisition, we present NMSE performance for the proposed SA-BCE and SA-WBCE method with different $N_{\mathrm{d}}$ in Fig. \ref{fig:CENMSEVerusSNRDifferentNd}. As shown in the figure, the performance of the proposed algorithm improves with the increasing $\text{SNR}_\mathrm{CE}$. Furthermore, the performance of the proposed algorithm improves as $N_{\mathrm{d}}$ increases. When \( N_{\mathrm{d}} \) is small, the condition in \eqref{VandermondeUniquenessCondition1} may not be satisfied, leading to inaccurate SCSI and degraded channel estimation performance. 
When \( N_{\mathrm{d}} \) is large, increasing \( N_{\mathrm{d}} \) does not significantly improve the channel estimation performance. Therefore, during the construction of the SCSI database, only a subset of measurement data is required. Furthermore, the SA-WBCE exhibits inferior performance compared to SA-BCE, particularly in the high-SNR regime. This result stems from the approximations made during the design process, which, in turn, contribute to a significant reduction in computational complexity.

\begin{figure}[!t]
	\centering
	\includegraphics[width=0.485\textwidth]{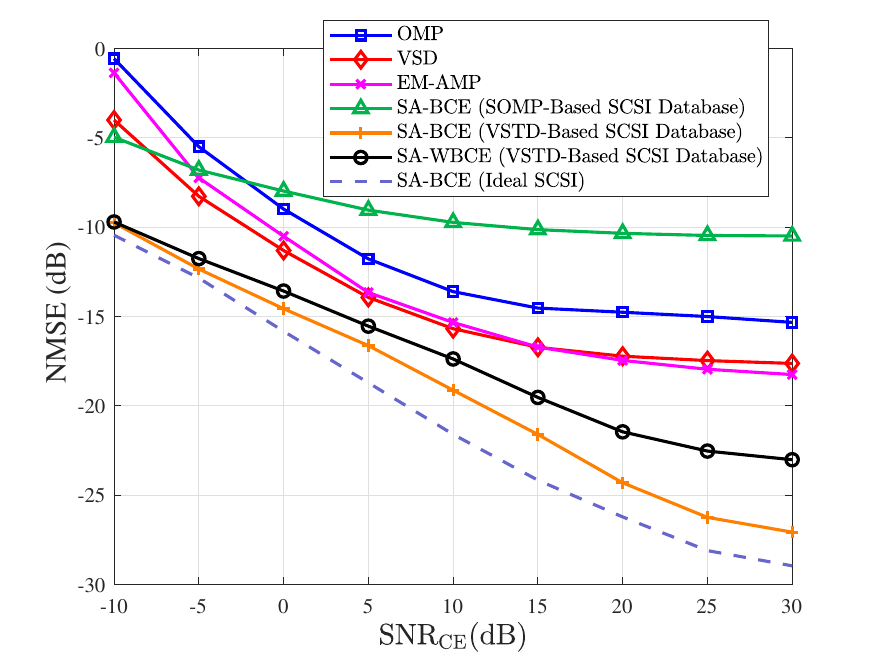}
	\caption{The NMSE of channel estimation versus SNR ($N_{\mathrm{d}}=240$, $\text{SNR}_\mathrm{SC}=10$ dB  and $d=2$m in SCSI database construction). }
	\label{fig:NMSECESNR}
\end{figure}

\begin{figure}[!t]
	\centering
	\includegraphics[width=0.485\textwidth]{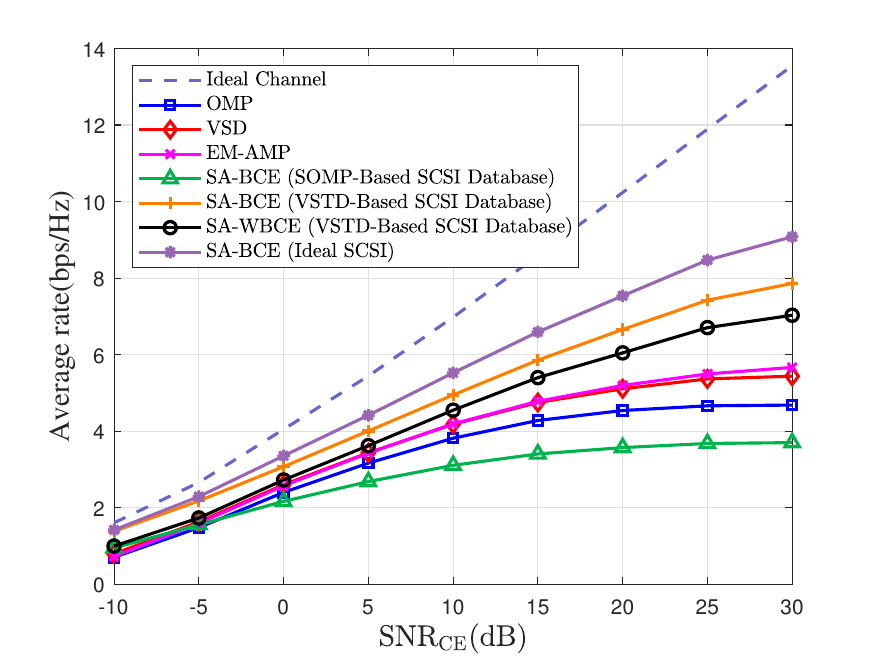}
	\caption{Average effective communication rate performance versus SNR ($N_{\mathrm{d}}=240$, $\text{SNR}_\mathrm{SC}=10$dB  and $d=2$m in SCSI database construction). }
	\label{fig:sim_AverageRate_SNR}
\end{figure}

The comparison of NMSE and average effective communication rate performance between the proposed schemes and baseline methods are shown in Fig. \ref{fig:NMSECESNR} and Fig. \ref{fig:sim_AverageRate_SNR}, respectively.  By leveraging the SCSI, the proposed schemes outperform all benchmark methods across the entire SNR range and achieves performance comparable to the SA-BCE with the ideal SCSI database.  In particular, the proposed SA-BCE and SA-WBCE with the VSTD-based SCSI database  achieve an NMSE of approximately $-21.5$ dB and $-19.5$ dB when $\text{SNR}_\mathrm{CE}$ is $15$ dB, whereas the NMSE of all other benchmark methods remains above $-17$ dB. 
These results demonstrate the great potential of the proposed SA-BCE and SA-WBCE schemes for MU-MIMO systems.

\begin{figure}[!t]
	\centering
	\includegraphics[width=0.485\textwidth]{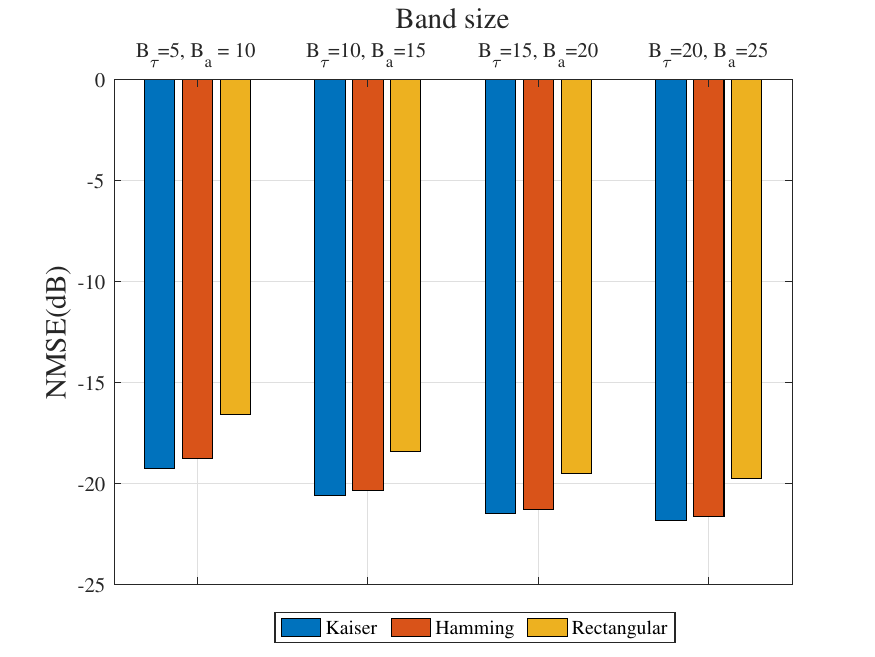}
	\caption{The NMSE performance of the SA-WBCE with different band size and different window functions  ($N_{\mathrm{d}}=240$, $\text{SNR}_\mathrm{SC}=10$dB  and $d=2$m in SCSI database construction, $\text{SNR}_\mathrm{CE}=20$dB). }
	\label{fig:sim_NMSE_TSWBCE_DifferentBandSize}
\end{figure}

Fig. \ref{fig:sim_NMSE_TSWBCE_DifferentBandSize} illustrates the NMSE performance of the SA-WBCE with the VSTD-based SCSI database under different band sizes and window functions. As the band size increases, the NMSE performance of the SA-WBCE scheme improves, albeit at the cost of higher computational complexity. In practical applications, a trade-off between performance and complexity can be achieved by appropriately selecting the band size. Among the three window functions, the Kaiser window offers the best performance. Specifically, When $B_{\tau} = 15$ and $B_{a} = 20$, the SA-WBCE with the Kaiser window achieves an NMSE of $–21.4$ dB,  representing a $2$ dB improvement than SA-WBCE with the rectangular window.

\begin{figure}[!t]
	\centering
	\includegraphics[width=0.485\textwidth]{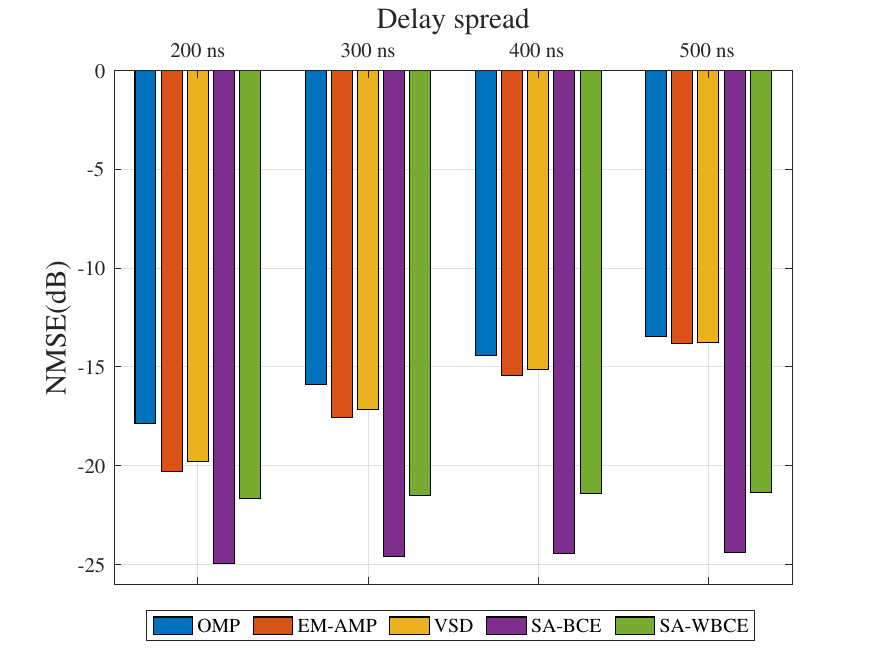}
	\caption{The NMSE performance of several schemes versus the delay spread   ($N_{\mathrm{d}}=240$, $\text{SNR}_\mathrm{SC}=10$dB  and $d=2$m in SCSI database construction, $\text{SNR}_\mathrm{CE}=20$dB). For the SA-WBCE scheme, the band sizes are set to $B_{\tau} = 15$ and $B_{a} = 20$, with the Kaiser window applied. }
	\label{fig:NmseCeVersusDelaySpread}
\end{figure}

In Fig. \ref{fig:NmseCeVersusDelaySpread}, we present the NMSE performance of various algorithms as a function of the channel delay spread.  The OMP, VSD, and EM-AMP algorithms, which rely on OCC decomposition results obtained through trivial approach for channel estimation, suffer significant performance degradation as the delay spread increases. In contrast, the proposed SA-BCE and SA-WBCE, which incorporate SCSI into the OCC decomposition process to mitigate pilot interference, exhibit minimal NMSE deterioration. Specifically, as the delay spread increases from $200$ ns to $500$ ns, the proposed algorithms experiences only a $0.5$ dB NMSE loss, compared to a $6.5$ dB degradation observed in the VSD and EM-AMP algorithms.

\section{Conclusion}
\label{sec_Conclusion}

In this paper, we investigated uplink DMRS-based channel estimation for MU-MIMO systems. We first developed the received signal model under the Type II OCC pattern standardized  in 3GPP Release 18. Building on this model, we proposed SA-BCE, which effectively suppresses pilot interference by fully leveraging SCSI. To further reduce the computational complexity of SA-BCE, we reformulated the estimation process from the antenna-frequency domain to the beam-delay domain and extended the approach to SA-WBCE by incorporating the windowing technique.
To acquire SCSI, we constructed a location-specific SCSI database by partitioning the spatial region into grids and leveraged the uplink received signals within each grid to extract the SCSI. Facilitated by the multilinear structure of wireless channels, we formulated  the SCSI acquisition problem within each grid as a tensor decomposition problem and exploited the VSTD algorithm to extract the SCSI.
Simulation results validated the superiority of the proposed schemes.

\normalem

\bibliographystyle{IEEEtran}
\bibliography{DMRS_Ref}

\end{document}